\documentclass[12pt]{article}
\usepackage{epsfig,amsmath,amsfonts,amssymb,amstext,afterpage,psfrag,
}
\usepackage{slashed}
\usepackage[square,comma,sort&compress,numbers]{natbib}
\allowdisplaybreaks
\def\l{\langle}
\def\r{\rangle}

\begin{document}

\begin{titlepage}

\begin{flushright}
{\small
LMU-ASC~69/15\\ 
October 2015
}
\end{flushright}

\vspace{0.5cm}
\begin{center}
{\Large\bf \boldmath                                               
Fitting Higgs Data with Nonlinear Effective Theory 
\unboldmath}
\end{center}

\vspace{0.5cm}
\begin{center}
{\sc G.~Buchalla, O.~Cat\`a, A.~Celis and C.~Krause} 
\end{center}

\vspace*{0.4cm}

\begin{center}
Ludwig-Maximilians-Universit\"at M\"unchen, Fakult\"at f\"ur Physik,\\
Arnold Sommerfeld Center for Theoretical Physics, 
D--80333 M\"unchen, Germany
\end{center}

\vspace{1.5cm}
\begin{abstract}
\vspace{0.2cm}
\noindent
In a recent paper we showed that the electroweak chiral Lagrangian at 
leading order is equivalent to the conventional $\kappa$ formalism used by 
ATLAS and CMS to test Higgs anomalous couplings. Here we apply 
this fact to fit the latest Higgs data. The new aspect of our analysis is a 
systematic interpretation of the fit parameters within an EFT. Concentrating 
on the processes of Higgs production and decay that have been measured so far, 
six parameters turn out to be relevant:
$c_V$, $c_t$, $c_b$, $c_\tau$, $c_{\gamma\gamma}$, $c_{gg}$.
A global Bayesian fit is then performed with the result:
$c_{V} = 0.98 \pm 0.09$, $c_{t} = 1.34 \pm 0.19$, $c_{b} = 0.78  \pm 0.18$,
$c_{\tau} = 0.92 \pm 0.14$, $c_{\gamma\gamma} = -0.24  \pm 0.37$,
$c_{gg} = -0.30  \pm 0.17$. 
Additionally, we show how this leading-order parametrization can be generalized 
to next-to-leading order, thus improving the $\kappa$ formalism systematically. 
The differences with a linear EFT analysis including operators of dimension 
six are also discussed. One of the main conclusions of our analysis is that 
since the conventional $\kappa$ formalism can be properly justified within a 
QFT framework, it should continue to play a central role in analyzing and 
interpreting Higgs data. 
\end{abstract}

\vfill

\end{titlepage}

\section{Introduction}
\label{sec:intro}

The first run of the LHC has witnessed the discovery of a Higgs-like particle 
and the determination of its prominent couplings with a typical precision
of 10-20\%, with no significant 
deviations from the Standard Model (SM). The main tool to measure Higgs 
couplings at Run 1 has been the so-called $\kappa$ formalism, a 
signal-strength parametrization at the level of the decay rates
and production cross sections. 
The $\kappa$ formalism was intended as a first tool to capture large 
deviations from the SM, and expected to be superseded by a more refined, 
QFT-based approach. One of the main goals of the second run is to increase 
the precision to the 5\% level and explore shapes. 
In this context an upgrade of the $\kappa$ formalism appears necessary.

In a recent paper~\cite{Buchalla:2015wfa} we have shown that the $\kappa$ 
formalism is actually the natural outcome of the nonlinear 
effective field theory (EFT) at leading order (LO). 
In order to make the EFT connection more transparent one needs to 
trade parameters that are nonlocal at the electroweak scale for those that 
are local. This applies in particular to loop-induced processes like 
$h\to\gamma\gamma$ and $h\to Z\gamma$. Since Lagrangian parameters are local, 
it is clear that this should help interpret the experimental results within 
theoretical frameworks. In particular, this choice of parameters also 
facilitates the incorporation of radiative corrections.

In this paper we will illustrate this aspect of the LO nonlinear EFT by 
performing a fit to the latest Higgs data. 
Compared to previous fits \cite{Carmi:2012yp,Azatov:2012bz}, 
our emphasis here is on the systematics: an EFT-based framework allows us 
to use Bayesian methods with priors supported by power-counting arguments, 
thereby allowing a consistent implementation of model-independent dynamical 
information into the fitting procedure.   

The fact that a signal-strength analysis of Higgs decays can be embedded into 
an EFT framework means that it is possible to go to next-to-leading order 
(NLO) in the expansion. 
This can be seen as the natural extension of the $\kappa$ formalism, where 
now parameters have to be defined at the amplitude level. Interestingly, what 
one finds is that NLO operators contribute to the shapes, thus making our EFT 
formalism suitable for analyses of Run 2 data and beyond. 
The set of parameters needed 
to go to NLO experimentally is discussed in Section~\ref{sec:NLO}. However, 
one of the conclusions of the EFT analysis is that deviations from the SM in 
the shapes are suppressed by roughly two orders of magnitude with respect to 
those in the rates. Thus, if the present 
10-20\% uncertainty in the rates turns out to hide NP effects of similar
size, the same dynamics 
will affect shapes only at the per-mille level, well out of the scope of the 
LHC even in its final stage.  
 
This paper will be organized as follows: in Section~\ref{sec:leff} we will 
spell out the structure of the chiral Lagrangian together with its 
underlying dynamical assumptions. The set of leading order parameters relevant 
to Higgs decays are fit to Run 1 data in Section~\ref{sec:fit}. 
In Section~\ref{sec:NLO} we discuss how the analysis should be extended to 
NLO together with a comparison between the linear and nonlinear realizations.  
Conclusions are given in Section~\ref{sec:concl} while technical aspects of 
the fitting procedure are relegated to an Appendix.

\section{Effective Lagrangian}
\label{sec:leff}

In \cite{Buchalla:2015wfa} we proposed a parametrization of anomalous
Higgs-boson couplings based on the leading-order electroweak chiral 
Lagrangian 
\cite{Feruglio:1992wf,Buchalla:2013eza,Buchalla:2012qq,Buchalla:2013rka}.
It is an important aspect of this parametrization that it 
provides us with a consistent EFT justification of the usual 
$\kappa$ formalism \cite{Heinemeyer:2013tqa}.

The starting point of a systematic derivation is the effective Lagrangian at 
leading order, which can be written as \cite{Buchalla:2013rka} 
\begin{eqnarray}\label{l2}
{\cal L}_2 &=& -\frac{1}{2} \langle G_{\mu\nu}G^{\mu\nu}\rangle
-\frac{1}{2}\langle W_{\mu\nu}W^{\mu\nu}\rangle 
-\frac{1}{4} B_{\mu\nu}B^{\mu\nu}
+\bar q i\!\not\!\! Dq +\bar l i\!\not\!\! Dl
 +\bar u i\!\not\!\! Du +\bar d i\!\not\!\! Dd +\bar e i\!\not\!\! De 
\nonumber\\
&& +\frac{v^2}{4}\ \l D_\mu U^\dagger D^\mu U\r\, \left( 1+F_U(h)\right)
+\frac{1}{2} \partial_\mu h \partial^\mu h - V(h) \nonumber\\
&& - v \left[ \bar q \left( Y_u +
       \sum^\infty_{n=1} Y^{(n)}_u \left(\frac{h}{v}\right)^n \right) U P_+r 
+ \bar q \left( Y_d + 
     \sum^\infty_{n=1} Y^{(n)}_d \left(\frac{h}{v}\right)^n \right) U P_-r
  \right. \nonumber\\ 
&& \quad\quad\left. + \bar l \left( Y_e +
   \sum^\infty_{n=1} Y^{(n)}_e \left(\frac{h}{v}\right)^n \right) U P_-\eta 
+ {\rm h.c.}\right]
\end{eqnarray}
with $U=\exp(2i\phi^a T^a/v)$ the Goldstone-boson matrix, 
$T^a$ the generators of $SU(2)$, and $P_\pm = 1/2\pm T_3$.
Here
\begin{equation}\label{fufv} 
F_U(h)=\sum^\infty_{n=1} f_{U,n} \left(\frac{h}{v}\right)^n, \qquad
V(h)=v^4\sum^\infty_{n=2} f_{V,n}\left(\frac{h}{v}\right)^n
\end{equation}
The right-handed quark and charged-lepton singlets are written as 
$u$, $d$, $e$. $q$ ($l$) denote the left-handed and $r$ ($\eta$) the 
right-handed quark (lepton) doublets. Generation indices have been suppressed.
The $Y_f$, $Y^{(n)}_f$ are matrices in generation space.   

\begin{figure*}[t]
\begin{center}
\includegraphics[width=14cm]{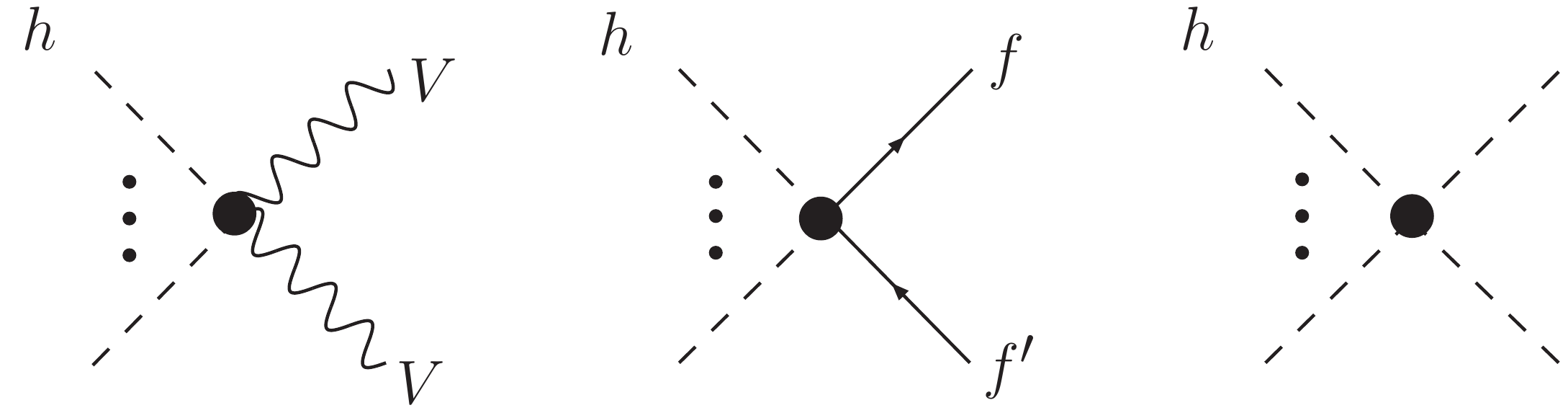}
\end{center}
\caption{\it The Higgs vertices from the leading-order Lagrangian ${\cal L}_2$
in unitary gauge.They are represented by a black dot and may deviate 
sizably from the SM.
The pair of dashed lines with dots in between signifies any number of 
Higgs lines. The massive vector bosons are denoted by $V=W,Z$.
$f=f'$ if flavour conservation is assumed to hold at leading order.
All other couplings are identical to the SM.}
\label{fig:hcouplings}
\end{figure*}

Let us summarize the essential properties of this Lagrangian:
\begin{itemize}
\item
The nonlinear EFT is organized in terms of a loop expansion or, equivalently,
in terms of chiral dimensions. The assignment of chiral dimensions 
is 0 for boson fields and 1 for derivatives, weak couplings 
and fermion bilinears \cite{Buchalla:2013eza}.
A chiral dimension of $2L+2$ for a term in the 
Lagrangian corresponds to loop order $L$. All the terms in (\ref{l2})
have a chiral dimension of 2. 
\item
The anomalous couplings $f_{U,n}$ and $f_{V,n}$ are, in general, 
arbitrary coefficients of order 1. They generalize the SM, 
in which the non-zero values are
\begin{equation}\label{fsm}
f_{U,1}=2,\quad f_{U,2}=1, \qquad f_{V,2}=f_{V,3}=\frac{m^2}{2 v^2},
\quad f_{V,4}=\frac{m^2}{8 v^2},
\end{equation}
where $m=125\,{\rm GeV}$ is the Higgs mass and $v=246\,{\rm GeV}$ the
electroweak vev.
If the relative deviations from the SM can be considered to be smaller 
than unity, it is convenient to parametrize them by a quantity
$\xi\equiv v^2/f^2 < 1$. $f$ corresponds to a new scale, which would represent
e.g. the Goldstone-boson decay constant in typical models of a composite 
Higgs \cite{Agashe:2004rs,Contino:2006qr,Contino:2010rs,Falkowski:2007hz,Carena:2014ria}.
From experiment, values of $\xi={\cal O}(10\%)$ are currently still allowed. 
A series expansion can be performed in $\xi$ if it is small enough.
This corresponds to an expansion of the effective theory in terms of
canonical dimensions. Using the coefficients of the chiral Lagrangian 
in (\ref{l2}) implies a resummation to all orders in $\xi$, at leading chiral
dimension.
Throughout this paper we will often call a deviation
from the SM to be of $\cal{O}(\xi)$ in the sense that it starts
at this order and understanding that all orders in $\xi$ are
included in the chiral Lagrangian coefficients.
An illustration of the systematics is provided in Fig. \ref{fig:eftplot}.
\begin{figure*}[t]
\begin{center}
\includegraphics[width=8cm]{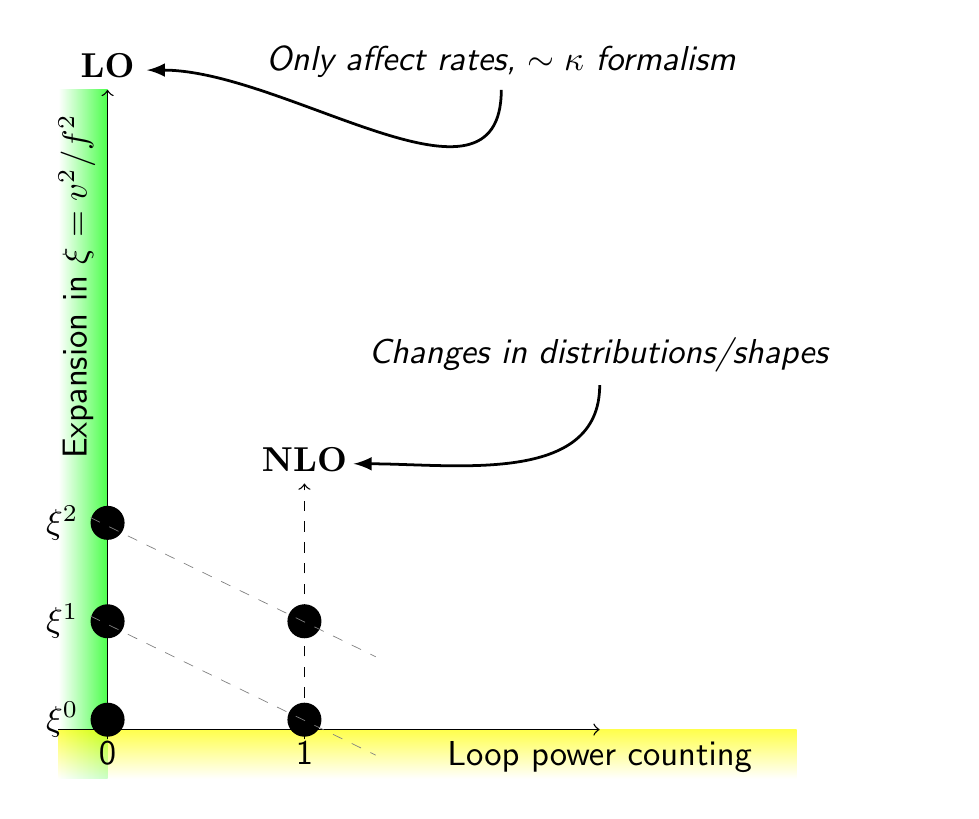}
\end{center}
\vspace*{-0.5cm}
\caption{\it Loop order vs. expansion in $\xi$.}
\label{fig:eftplot}
\end{figure*}
As mentioned above, for tree-level processes, deviations from the 
SM in the distributions arise at NLO and are suppressed by roughly two orders 
of magnitude with respect to the LO effects in the rates. 
It is important to stress that this is a dynamical feature, not a kinematical 
one: deviations in the 
shapes are suppressed not based on phase space considerations but merely 
as a prediction of the underlying dynamics of the EFT. In other words, the 
nonlinear EFT dynamically separates rates as LO-sensitive and shapes as 
NLO-sensitive observables.\footnote{An important qualification of this
generic statement is discussed at the end of this section.} 
This is unlike the linear EFT, where deviations 
from the SM in rates and shapes are both expected at the few-percent level.
\item
The Yukawa couplings $Y_f$, $Y^{(n)}_f$ may formally all be considered to be
of order unity as far as the chiral counting is concerned.
This is realistic only for the top quark. The other Yukawa couplings
come with a strong numerical suppression from flavour physics,
{\it a priori} unrelated to chiral counting. As usual, this suppression 
can be used to make corresponding approximations in applications.
Note that since $Y^{(1)}$ is in general independent of $Y$, flavour-changing 
couplings of the Higgs to fermions can naturally be accommodated by 
(\ref{l2}).
In the SM one has $Y^{(1)}_f=Y_f$, corresponding to the usual Yukawa matrices,
while the remaining $Y^{(n)}_f$ are zero. Similar to the discussion in the
previous item, deviations from the SM can be described by the parameter $\xi$.
\item
In writing (\ref{l2}) we have assumed that custodial symmetry is respected
by the (strong) dynamics underlying the Higgs sector, and is only violated
by weak perturbations. Such perturbations then come with a weak coupling,
e.g. from gauge or Yukawa interactions, 
which carries chiral dimension \cite{Buchalla:2014eca}.
The operators violating custodial symmetry are then shifted to higher order
in the chiral expansion. For instance, the operator 
\begin{equation}\label{beta1}
v^2 \lambda^2_c \langle T_3 U D_\mu U^\dagger \rangle^2 (1+F_{\beta_1}(h))
\end{equation}
(related to the electroweak $T$-parameter)
breaks custodial symmetry due to the presence of $T_3$ under the trace.
If $T_3$ is associated with a weak coupling $\lambda_c$, the chiral dimension 
of (\ref{beta1}) is four in total, corresponding to a next-to-leading
order effect. For this reason the two-derivative operator in (\ref{beta1})     
does not have to be included in (\ref{l2}).
\item
The leading-order Lagrangian (\ref{l2}) consistently describes anomalous
Higgs interactions, with potentially sizable deviations from the SM.
By contrast, the gauge interactions are exactly as in the SM at this order.
A pictorial summary of the general Higgs couplings contained in (\ref{l2})
is given in Fig. \ref{fig:hcouplings}.
\end{itemize}

A special consideration is required for the application of the chiral
Lagrangian to processes that arise only at one-loop level in the SM.
Important examples are $h\to gg$ and $h\to\gamma\gamma$.
In this case local terms at NLO will also become
relevant, in addition to the standard loop amplitudes with modified 
couplings from (\ref{l2}). This is because those terms can lead to
deviations of the amplitude from the SM at the same order, $\sim \xi/16\pi^2$.

There is exactly one CP-even NLO operator contributing a local $h\to gg$
vertex,
\begin{equation}\label{xh3}
{\cal O}_{Xh3}=g^2_s \langle G_{\mu\nu}G^{\mu\nu}\rangle\, F_{Xh3}(h)
\end{equation}
in the notation of \cite{Buchalla:2013rka}.

For $h\to\gamma\gamma$ the following three operators from the complete
basis in \cite{Buchalla:2013rka} are relevant
\begin{eqnarray}\label{xhi}
{\cal O}_{Xh1} &=& g'^2 B_{\mu\nu} B^{\mu\nu}\, F_{Xh1}(h)\nonumber\\
{\cal O}_{Xh2} &=& g^{2} \langle W_{\mu\nu} W^{\mu\nu}\rangle\, F_{Xh2}(h)
\nonumber\\
{\cal O}_{XU1} &=& g' g B_{\mu\nu}\langle W^{\mu\nu} U T_3 U^\dagger\rangle\, 
(1+F_{XU1}(h))
\end{eqnarray}
They induce four couplings of a single Higgs to a pair of gauge bosons,
which in the physical basis with photon ($F_{\mu\nu}$), $Z$-boson ($Z_{\mu\nu}$)
and charged $W$ ($W^\pm_{\mu\nu}$) fields are given by
\begin{equation}\label{hxx}
e^2F_{\mu\nu}F^{\mu\nu}h,\qquad e g' F_{\mu\nu}Z^{\mu\nu}h;\qquad
g'^2 Z_{\mu\nu}Z^{\mu\nu}h,\qquad g^2 W^+_{\mu\nu} W^{-\mu\nu}h
\end{equation}
Since the four terms in (\ref{hxx}) arise from only three independent
operators (\ref{xhi}), their four coefficients are related
(see (\ref{cirel})).

The first two terms in (\ref{hxx}) give leading contributions to the 
loop-induced processes $h\to\gamma\gamma$ and $h\to Z\gamma$, respectively,
and have to be retained in a LO analysis. On the other hand, the last two
terms yield only subleading contributions, of ${\cal O}(\xi/16\pi^2)$,
to the tree-level amplitudes for $h\to ZZ$ and $h\to W^+W^-$, which receive
new-physics corrections of ${\cal O}(\xi)$ from (\ref{l2}). 
They can thus be neglected in a first approximation
(see Section 4 for the discussion of NLO effects). 

We add the following remarks:
\begin{itemize}
\item
In the full basis of the chiral Lagrangian at NLO \cite{Buchalla:2013rka}
a further operator 
\begin{equation}\label{xu2}
{\cal O}_{XU2} = g^2 \langle W_{\mu\nu}UT_3 U^\dagger\rangle^2 (1+F_{XU2}(h))
\end{equation}
could be written, in the same class as the operators in (\ref{xhi}).
However, this operator breaks custodial symmetry through the presence of
the generator $T_3$, which is unrelated to the factors of $W_{\mu\nu}$ and the 
associated coupling $g$. Since we assume that the breaking of custodial
symmetry through $T_3$ is due to weak perturbations, it has to come with
another weak coupling of chiral dimension one. The operator then acquires in 
total a chiral dimension of six, and is subleading to the terms in (\ref{xhi}).
\item
CP-odd structures corresponding to (\ref{xh3}) and (\ref{xhi}) of the type
$\varepsilon^{\mu\nu\lambda\rho}\langle W_{\mu\nu} W_{\lambda\rho}\rangle$ 
are part of the complete basis
and could also be considered. We will assume that CP symmetry in the Higgs 
sector is only broken by weak interactions. The CP-odd terms are then of 
higher order in the EFT and can be consistently neglected.
It would be straightforward to relax this assumption and to take those
terms into account.
\end{itemize}

\begin{figure*}[t]
\begin{center}
\includegraphics[width=14cm]{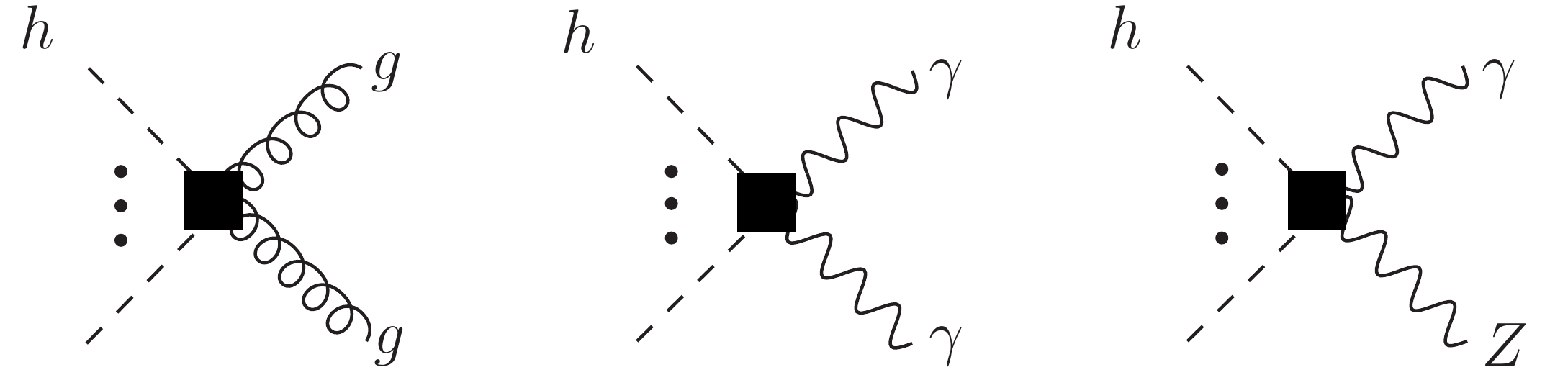}
\end{center}
\caption{\it Higgs vertices from the NLO Lagrangian ${\cal L}_4$, represented 
by black squares, that contribute to $gg$, $\gamma\gamma$ and $Z\gamma$ 
amplitudes. Since the latter arise only at one-loop order from the 
interactions of ${\cal L}_2$, the NLO couplings give relative corrections of 
the same order in this case and have to be retained.}
\label{fig:hnlo}
\end{figure*}

To summarize, the Higgs couplings from NLO operators that are relevant for a 
LO analysis of loop-induced processes are illustrated in Fig. \ref{fig:hnlo}.

\vspace*{0.4cm}

Based on the preceding discussion, we can now define anomalous Higgs
couplings for specific classes of interactions, corresponding to the
leading order approximation within the chiral Lagrangian framework.

An important example are interactions involving a single Higgs field.
Focusing on these terms, and working in unitary gauge, (\ref{l2}) 
supplemented by the local NLO terms for $h\to\gamma\gamma$, $Z\gamma$ and 
$gg$, implies  the interaction Lagrangian 
\begin{equation}\label{llopar1}
\begin{array}{ll}
  \mathcal{L} &=2 c_{V} \left(m_{W}^{2}W_{\mu}^{+}W^{-\mu} 
+\frac{1}{2} m^2_Z Z_{\mu}Z^{\mu}\right) \dfrac{h}{v} \vspace*{0.3cm} \\
&-\sum_{i,j} (y^{(1)}_{u,ij} \bar u_{Li}u_{Rj} + y^{(1)}_{d,ij} \bar d_{Li}d_{Rj}
  +y^{(1)}_{e,ij} \bar e_{Li}e_{Rj} + {\rm h.c.}) h \vspace*{0.3cm} \\
 &+ \dfrac{e^{2}}{16\pi^{2}} c_{\gamma\gamma} F_{\mu\nu}F^{\mu\nu} \dfrac{h}{v}  
+ \dfrac{e g'}{16\pi^{2}} c_{Z\gamma} Z_{\mu\nu}F^{\mu\nu} \dfrac{h}{v}
+\dfrac{g_{s}^{2}}{16\pi^{2}} c_{gg}\langle G_{\mu\nu}G^{\mu\nu}\rangle\dfrac{h}{v}
\end{array}
\end{equation}
Neglecting flavour violation, the very small Yukawa couplings to
light fermions, and concentrating on those Higgs processes that have already
become accessible at the LHC, the parametrization reduces to
a simple set of six anomalous couplings, described by \cite{Buchalla:2015wfa}
\begin{equation}\label{llopar2}
\begin{array}{ll}
  \mathcal{L} &=2 c_{V} \left(m_{W}^{2}W_{\mu}^{+}W^{-\mu} 
+\frac{1}{2} m^2_Z Z_{\mu}Z^{\mu}\right) \dfrac{h}{v} -c_{t} y_{t} \bar{t} t h 
- c_{b} y_{b} \bar{b}b h -c_{\tau} y_{\tau} \bar{\tau}\tau h  \\
 &+ \dfrac{e^{2}}{16\pi^{2}} c_{\gamma\gamma} F_{\mu\nu}F^{\mu\nu} \dfrac{h}{v}  
+\dfrac{g_{s}^{2}}{16\pi^{2}} c_{gg}\langle G_{\mu\nu}G^{\mu\nu}\rangle\dfrac{h}{v}
\end{array}
\end{equation}
where $y_f=m_f/v$.
The SM at tree level is given by $c_{V}= c_{t} = c_{b}=c_{\tau} =1$ and 
$c_{gg} =c_{\gamma\gamma} =0$. Deviations due to new physics are expected to 
start at $\mathcal{O}(\xi)$.

The minimal version in (\ref{llopar2}) can be generalized to include
more of the couplings contained in (\ref{llopar1}), such as
$h\to Z\gamma$, $h\to\mu\mu$, or the lepton-flavour violating $h\to\tau\mu$.

The treatment can be further extended, for instance to double-Higgs
production, where additional couplings with two or three $h$-fields
from (\ref{l2}) need to be considered. 

We would like to emphasize an important aspect of the
nonlinear EFT at leading order.
The anomalous couplings $c_i$ are able to account for deviations of
${\cal O}(1)$ from the SM. It is then consistent to retain the terms
quadratic in these couplings when computing cross sections and rates.
This is in contrast to the linear case, where a linearization in the 
dimension-6 corrections has to be performed at this level of accuracy.

A final remark concerns the above-mentioned distinction between
LO coefficients, affecting the rates, and NLO terms, modifying
decay distributions. Such a correspondence holds for
tree-level induced reactions such as $h\to Zl^+l^-$. By contrast, 
loop-induced processes have the property to exhibit non-standard distributions
even at leading order in the chiral description. An interesting example
is the $p_T$-distribution of highly-boosted Higgs in gluon-gluon fusion.
As discussed in \cite{Grojean:2013nya}, this observable has the potential 
to yield important independent information on the coefficients $c_t$ and 
$c_{gg}$ in (\ref{llopar2}), while the inclusive $gg\to h$ rate only 
constrains their sum.


\section{Fitting the Higgs data}
\label{sec:fit}

We perform a global Bayesian inference analysis for the parameters 
$\{c_V, c_{t}, c_{b}, c_{\tau}, c_{\gamma \gamma}, c_{gg}\}$ 
defined in (\ref{llopar2}).    
We are interested in the posterior probability density function (pdf), 
which gives the conditional probability of the parameters, given the data.    
In Bayesian inference the posterior pdf is given by the normalized product of 
the likelihood (conditional probability of the data, given the parameters) 
and the priors~\cite{D'Agostini:2003qr}.    The publicly available code 
{\tt Lilith-1.1.3}~\cite{Bernon:2015hsa} is used to extract the likelihood 
from experimental results in which the production and decay modes have been 
unfolded from experimental categories. We take into account the latest 
determination of the Higgs signal strengths by the Tevatron and the LHC 
collaborations contained in the {\tt Lilith} database 
{\tt DB 15.09}~\cite{dbref}:
\begin{itemize}
\item ATLAS and CMS measurements of the Higgs boson production and decay 
rates using $\sqrt{s} =7$ and $8$~TeV data~\cite{combid,Aad:2015gra}, 
\cite{Khachatryan:2014jba,Khachatryan:2015ila,Aad:2014xzb,ATLAS007,Aad:2014eha,Khachatryan:2014qaa,multil},
considering the main Higgs decay channels: $bb$, $\tau\tau$, $\gamma \gamma$, 
$ZZ^*$ and $WW^{*}$.   
\item Measurement of the associated production rate 
$V H \rightarrow V b \bar b$ by the Tevatron~\cite{Aaltonen:2013kxa}.   
\end{itemize}
Deviations from the SM of $\mathcal{O}(10-20\%)$ are allowed in general by 
current Higgs data~\cite{combid,Aaltonen:2013kxa}, corresponding to a scale of 
the strong dynamics $f\sim 500$ -- $1000$~GeV. New physics contributions to 
the parameters $\{c_V, c_{t}, c_{b}, c_{\tau}, c_{\gamma \gamma}, c_{gg}\}$ are 
expected to be of order $\mathcal{O}(\xi)$ due to the general power-counting 
arguments discussed in the previous section. Bayesian inference methods allow 
us to incorporate this knowledge in a systematic way through the application 
of Bayes's theorem and an appropriate choice of priors. For our analysis we 
use flat priors within the ranges: 
$c_V \in [0.5,1.5]$, $c_{f=t,b,\tau} \in [0,2]$, $c_{\gamma \gamma}\in [-1.5,1.5]$ 
and $c_{gg}\in [-1,1]$.\footnote{Flat priors for the Higgs couplings have also 
been used in previous Bayesian analyses of Higgs data~\cite{Azatov:2012bz}.}  
These priors allow for deviations in the parameters $c_i$ to be as large as 
$\sim 10\times \mathcal{O(\xi)}$. At the same time, they exclude additional 
disconnected solutions involving very large deviations from the SM for some 
of the parameters $c_{i}$. The fact that Bayesian methods make the inherent 
ambiguity in defining priors explicit is a useful feature when analyzing 
Higgs data within EFT, rather than being a disadvantage. More sophisticated 
treatments of the priors in which the notion of $\mathcal{O}(\xi)$ is 
parametrized by nuisance parameters can be naturally implemented in the 
Bayesian 
framework~\cite{Schindler:2008fh,Wesolowski:2015fqa}, though this is beyond the scope of our work.

\begin{figure}[!ht]
\centering
\includegraphics[width=0.3\textwidth]{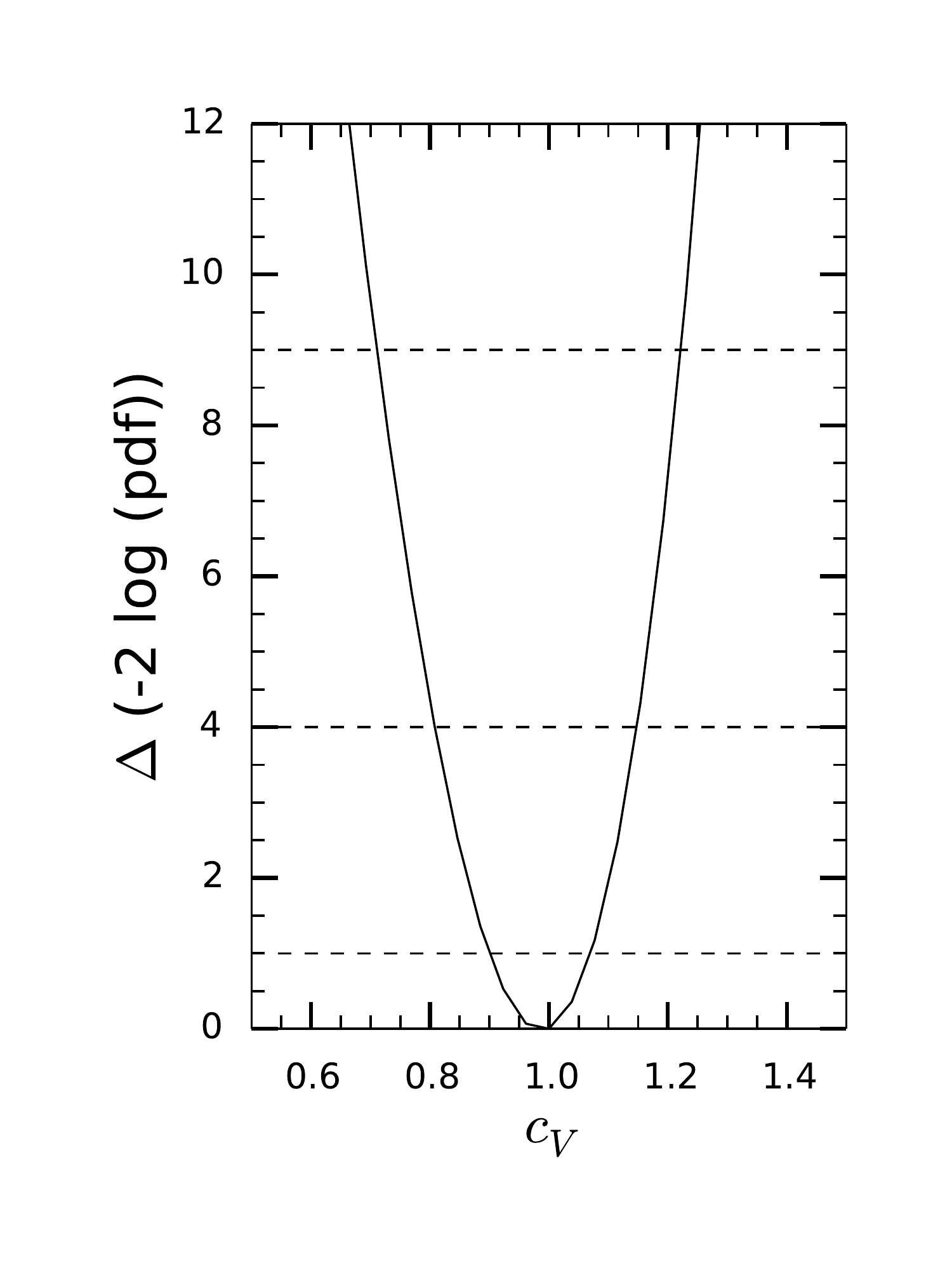} 
~  
\includegraphics[width=0.3\textwidth]{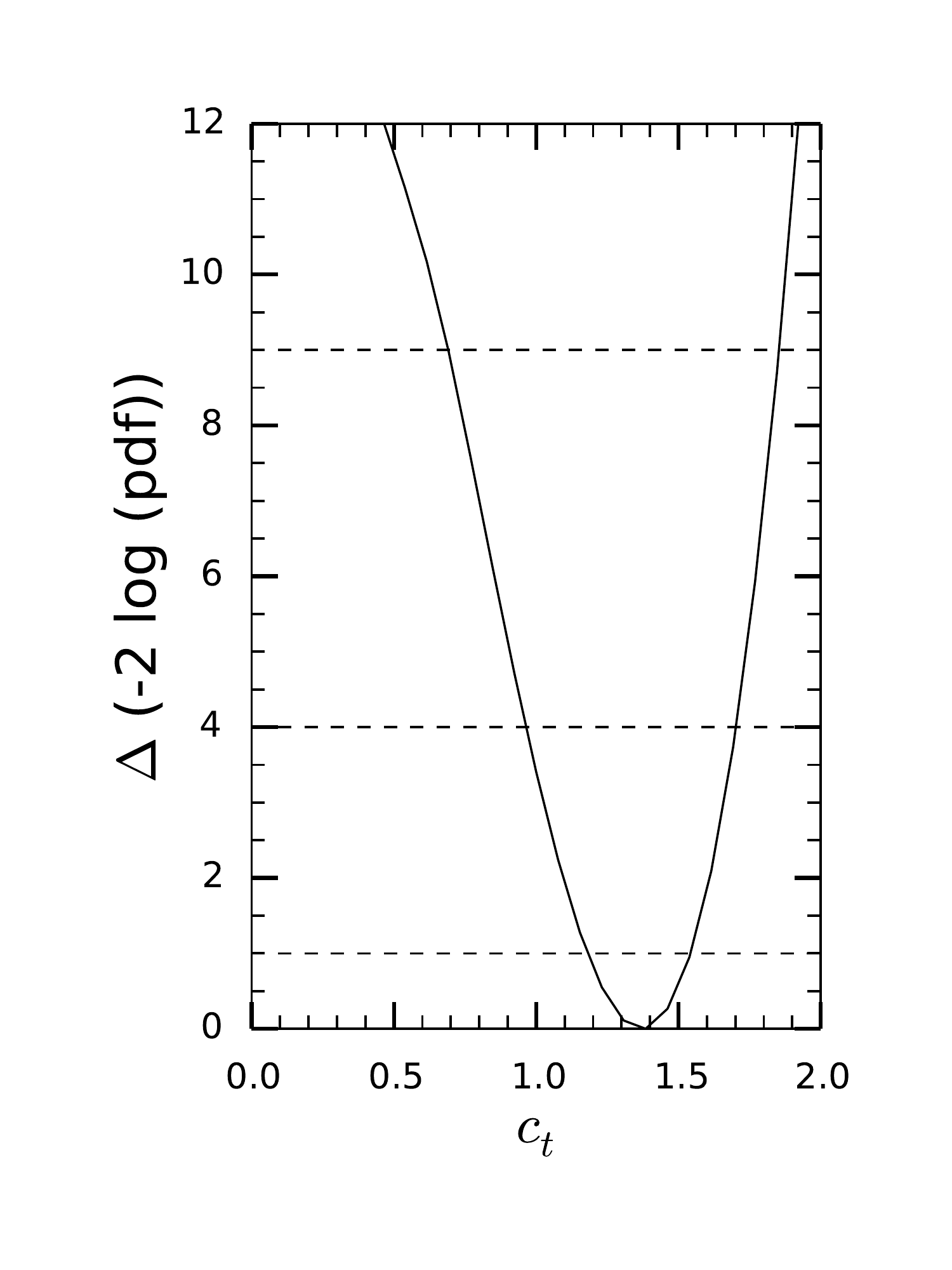}
~
\includegraphics[width=0.3\textwidth]{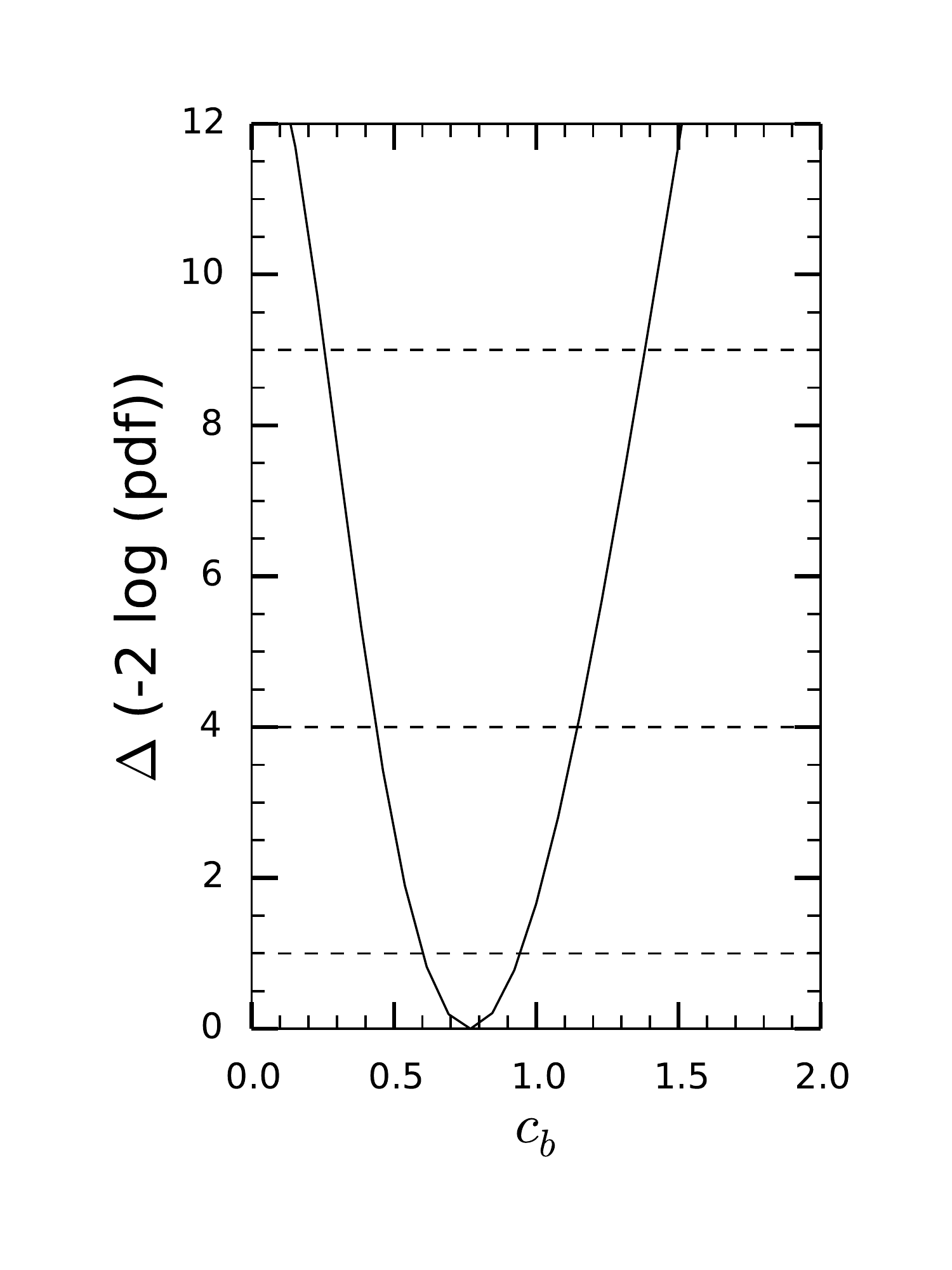}\\ \vspace{-0.3cm}
~
\includegraphics[width=0.3\textwidth]{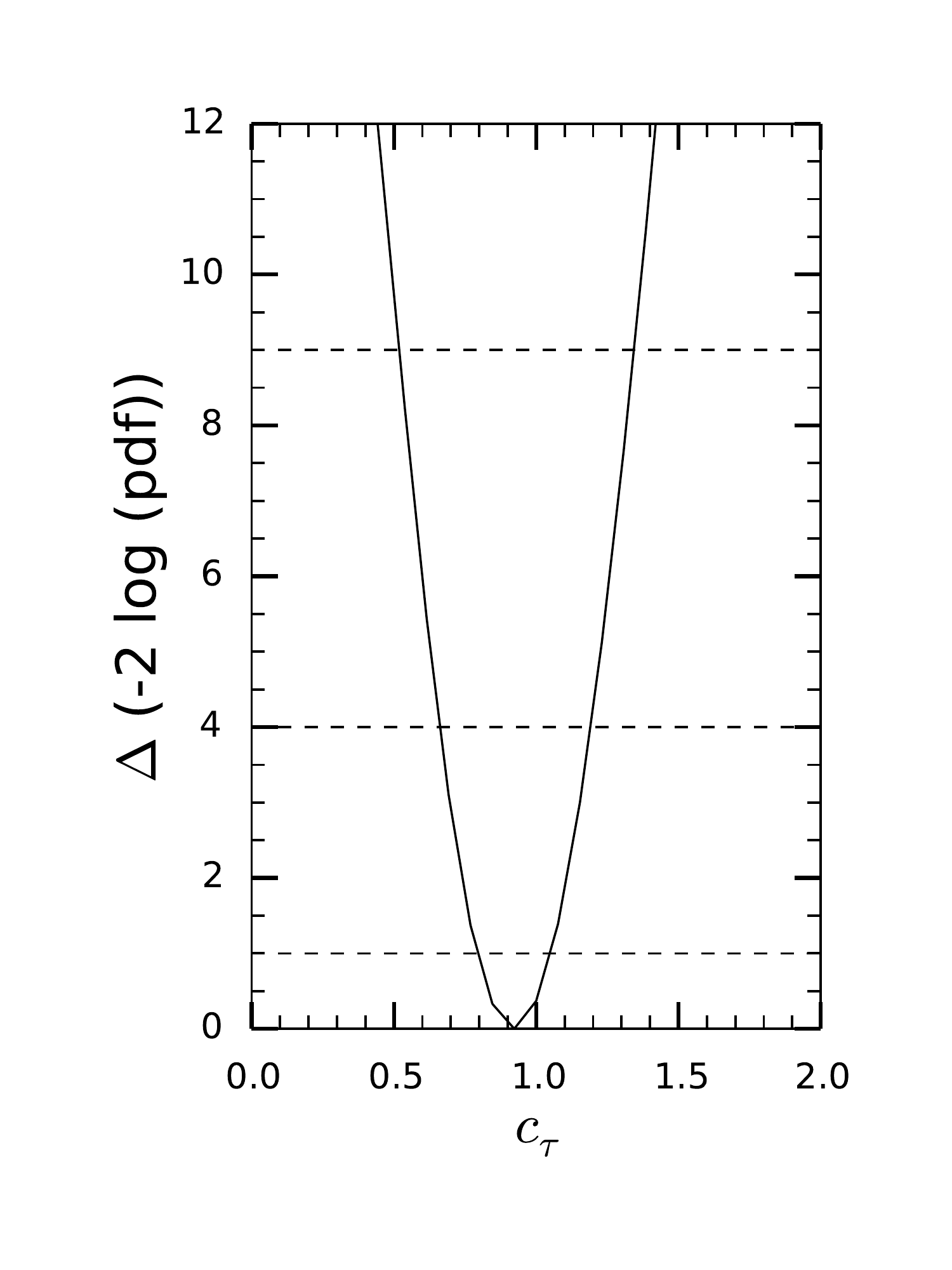}
~
\includegraphics[width=0.3\textwidth]{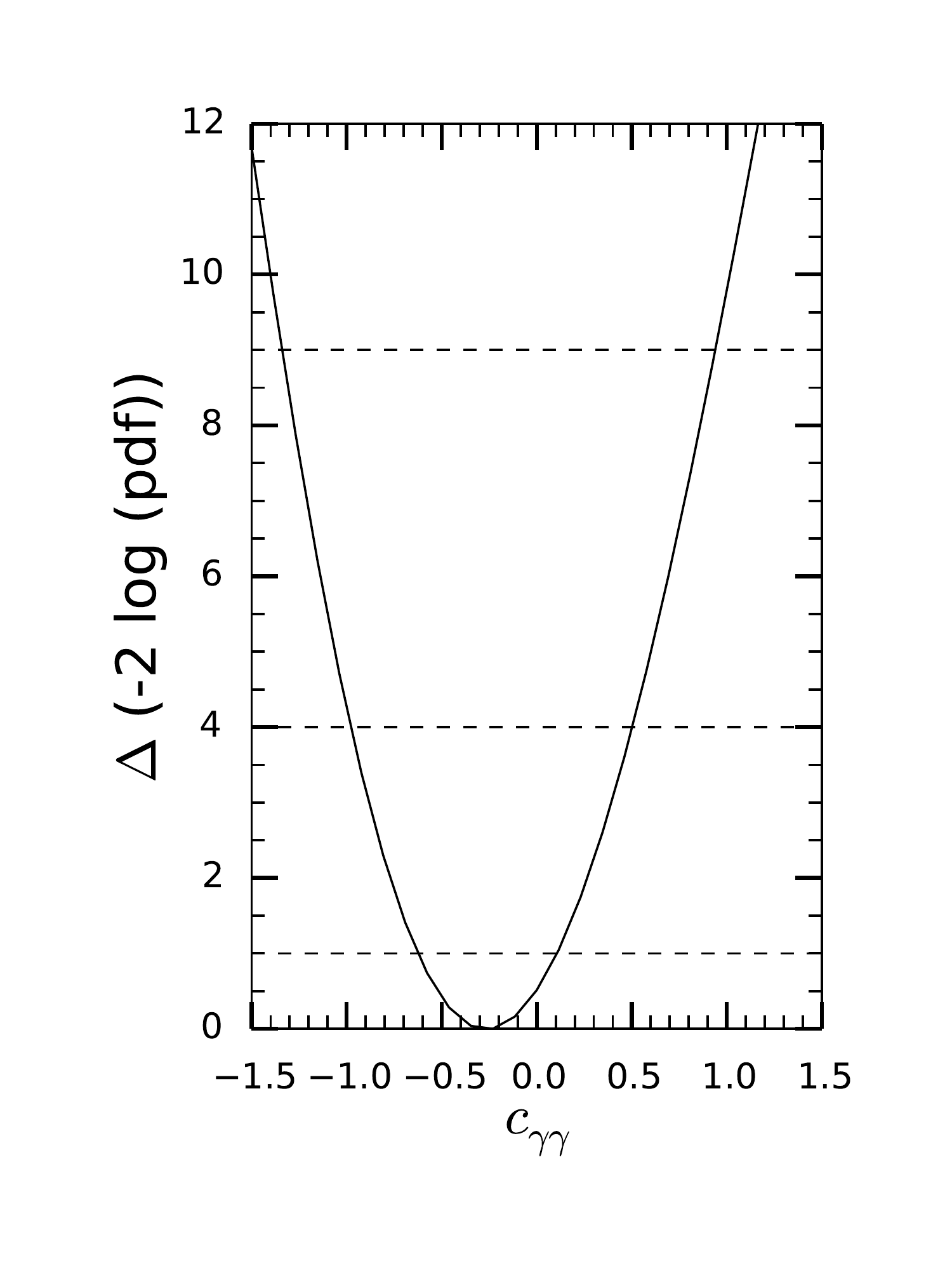}
~
\includegraphics[width=0.3\textwidth]{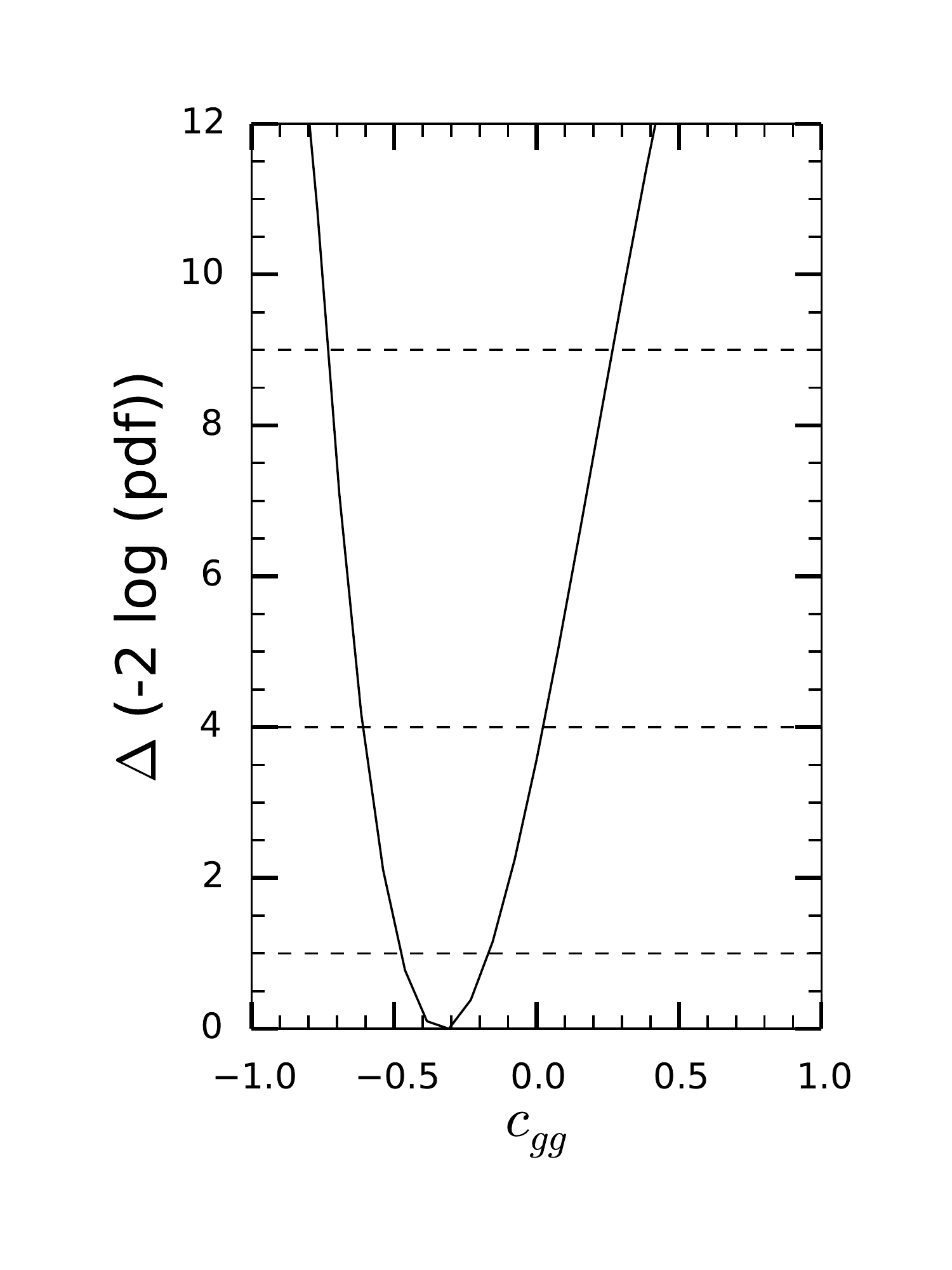}
\vspace{-0.6cm}
\caption{\label{1dplots} \it \small  $\Delta \chi^2 $ distribution for the 
one-dimensional marginalized posterior pdf.      }
\end{figure}
%

We find that the posterior pdf attains its maximum value at 
\begin{equation}\label{cifit}
\{c_V, c_{t}, c_{b}, c_{\tau}, c_{\gamma \gamma}, c_{gg}\} = 
\{0.96, 1.38, 0.69, 0.92, -0.35, -0.38\}
\end{equation}
In Figure~\ref{1dplots} we show the 
marginalized pdf for each of the parameters in (\ref{llopar2}). For 
convenience we plot $\Delta \chi^2 = \chi^2 - \chi^2_{\mbox{\scriptsize{min}}}$ 
with  $ \chi^2 \equiv -2   \log (\mbox{pdf})$.   
Since the posterior pdf is well approximated by a normal distribution around 
the maximum of the pdf, isocontours of $\Delta \chi^2= 1,4,9$, shown in 
Figure~\ref{1dplots} as dashed lines, correspond to $68\%, 95\%, 99.7\%$ 
Bayesian credible intervals to a very good approximation. 
The marginalized mean 
values and standard deviations obtained from the posterior pdf, together with 
the correlation matrix, are 
\begin{align}
  \label{eq:fitresult}
  \begin{pmatrix}
  c_{V}    \\
  c_{t}    \\
  c_{b}    \\
  c_{\tau}    \\
  c_{\gamma\gamma}   \\
  c_{gg} \end{pmatrix} = 
  \begin{pmatrix}
   0.98 \pm 0.09   \\
   1.34 \pm 0.19   \\
   0.78  \pm 0.18   \\
   0.92 \pm 0.14   \\
   -0.24  \pm 0.37   \\
   -0.30  \pm 0.17 \end{pmatrix}
     \qquad   \rho = \left( \begin{array}{rrrrrr}
    1.0 &  0.01 &  0.67 &  0.37 &  0.41 &  0.1   \\
      .   &  1.0   &  0.02 & -0.05  &-0.36 &-0.81   \\
      .   &    .     &  1.0   &  0.58 & 0.02 &  0.37   \\
      .   &    .     &     .    &  1.0   &-0.05 &  0.26   \\
      .   &    .     &     .    &    .     &  1.0   &  0.30   \\
      .   &    .     &     .    &    .     &    .     &  1.0     \\
  \end{array}\right) 
\end{align}\\
%
\begin{figure}[p]
\centering
\vspace{-0.5cm}
\includegraphics[width=0.4\textwidth]{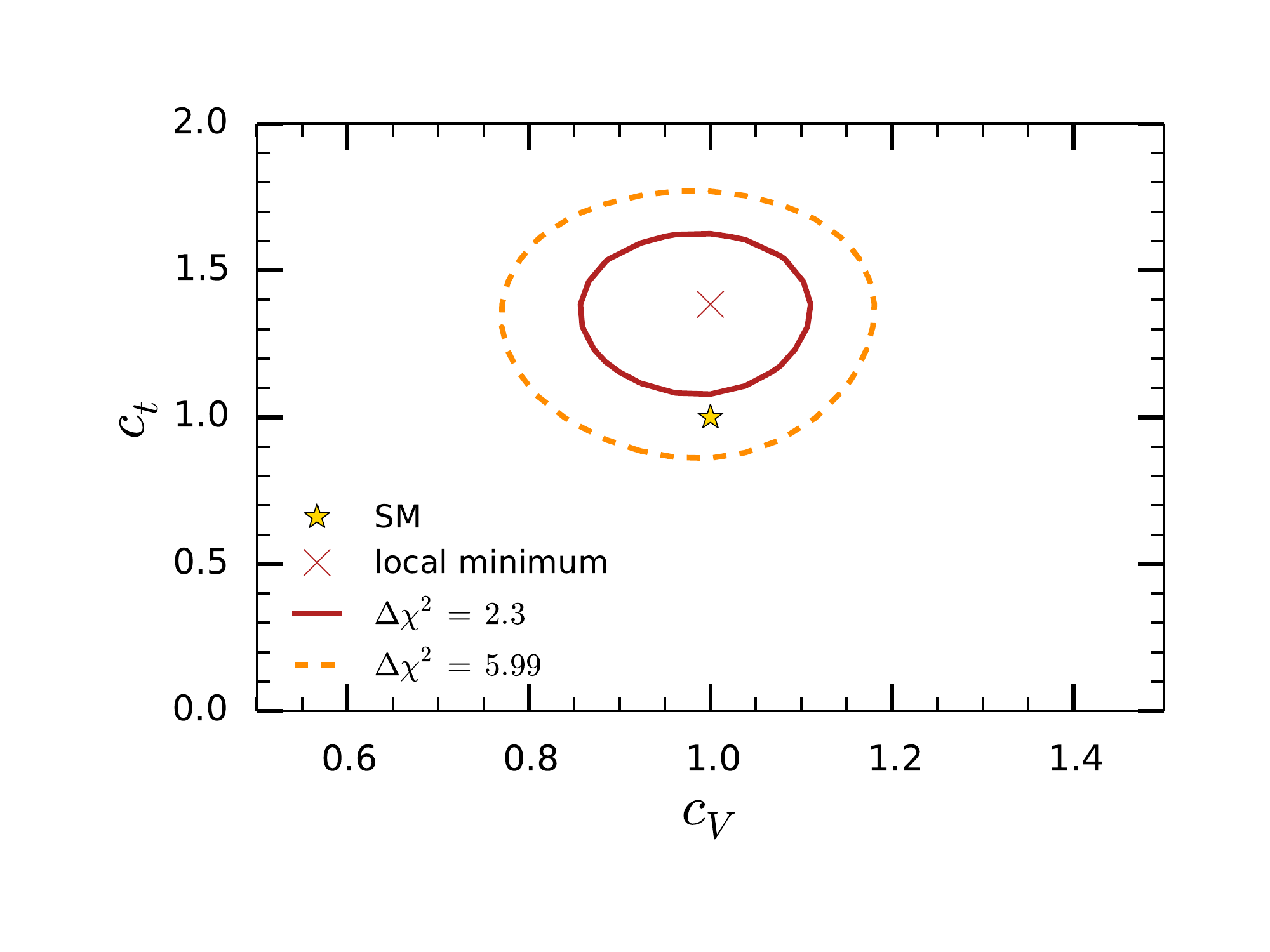}
~
\includegraphics[width = 0.4\textwidth]{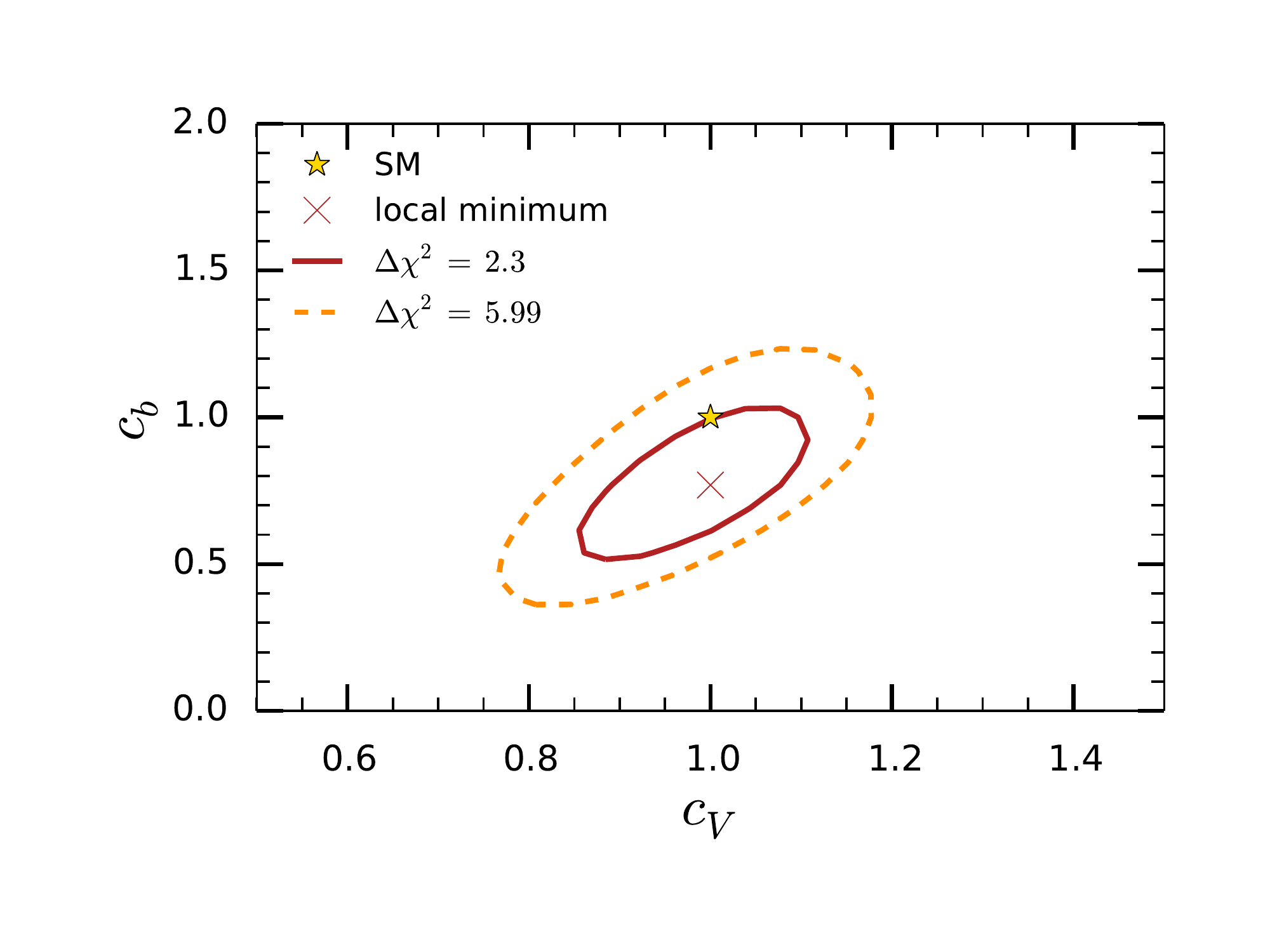}\\ \vspace{-0.6cm}
\includegraphics[width=0.4\textwidth]{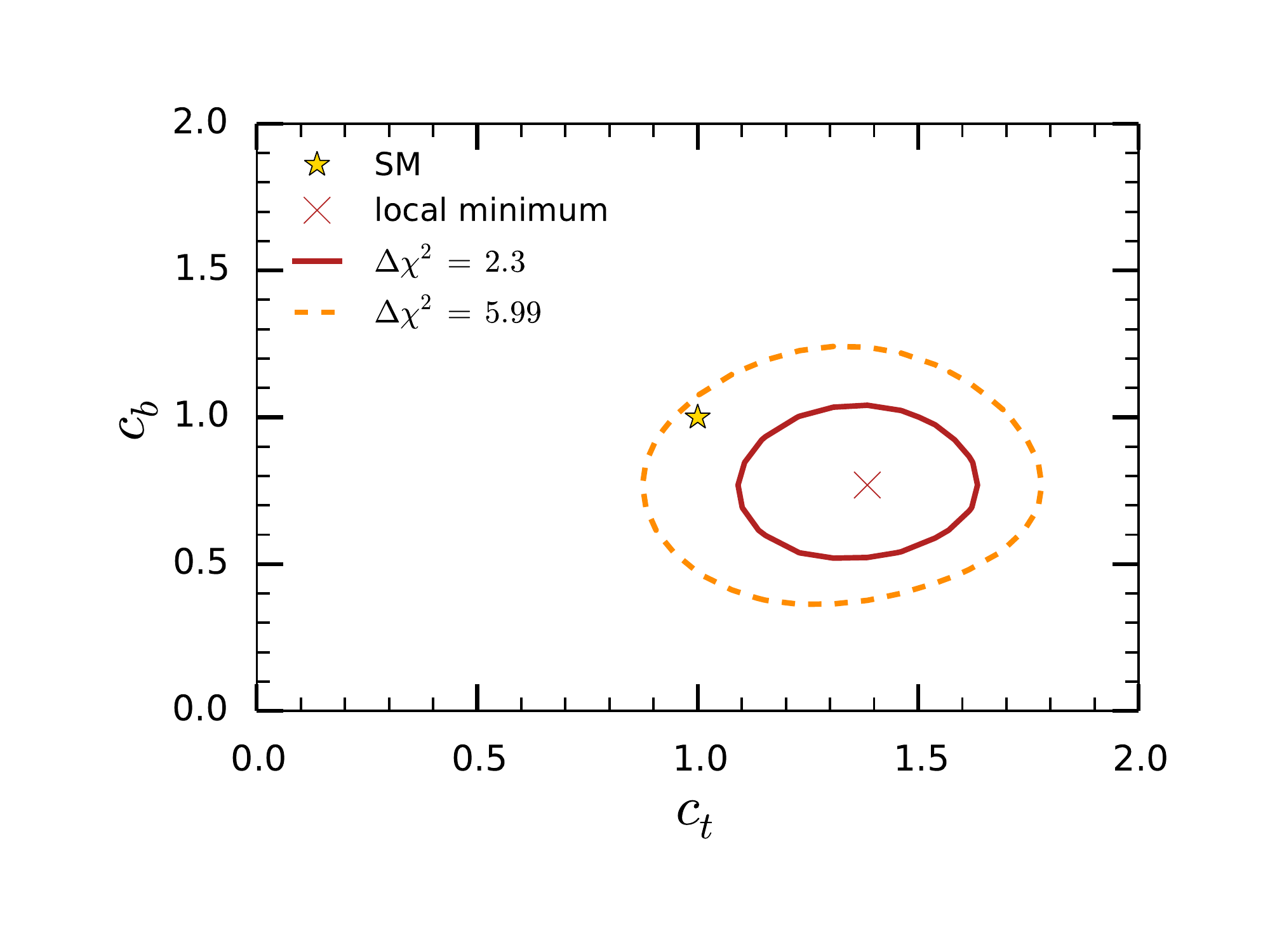}
~
\includegraphics[width = 0.4\textwidth]{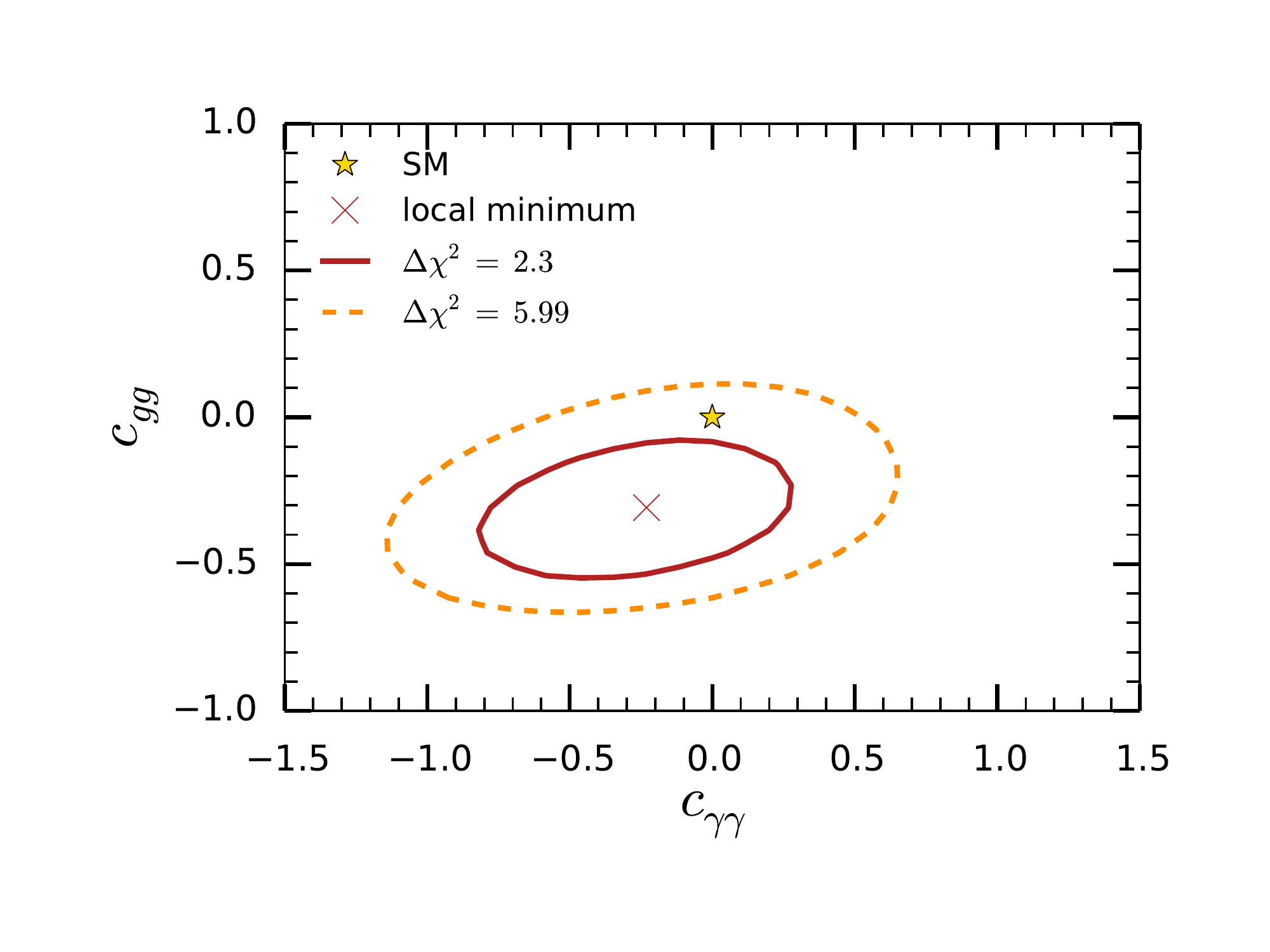}\\ 
\vspace{-0.6cm}
\includegraphics[width=0.4\textwidth]{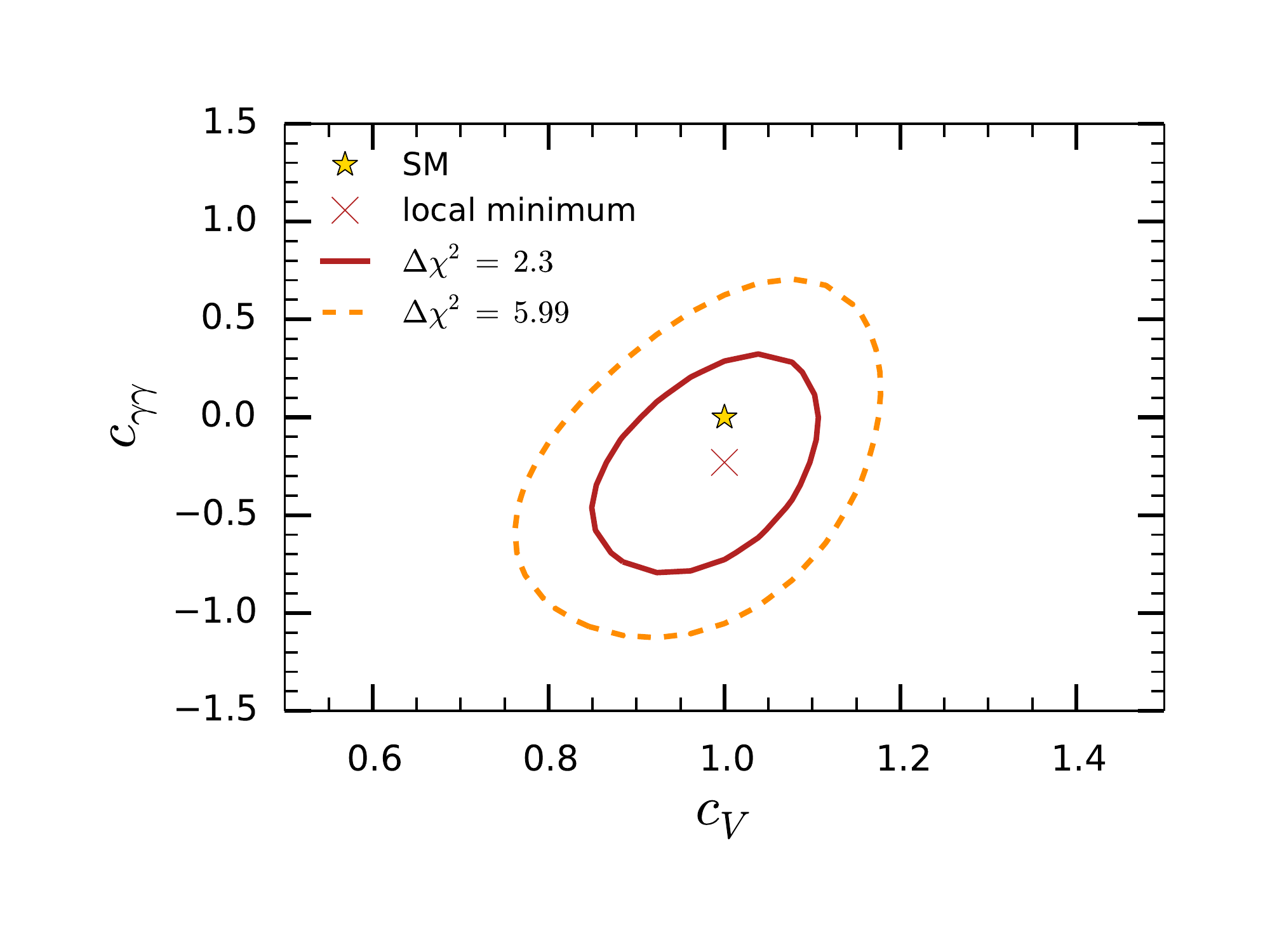}
~
\includegraphics[width = 0.4\textwidth]{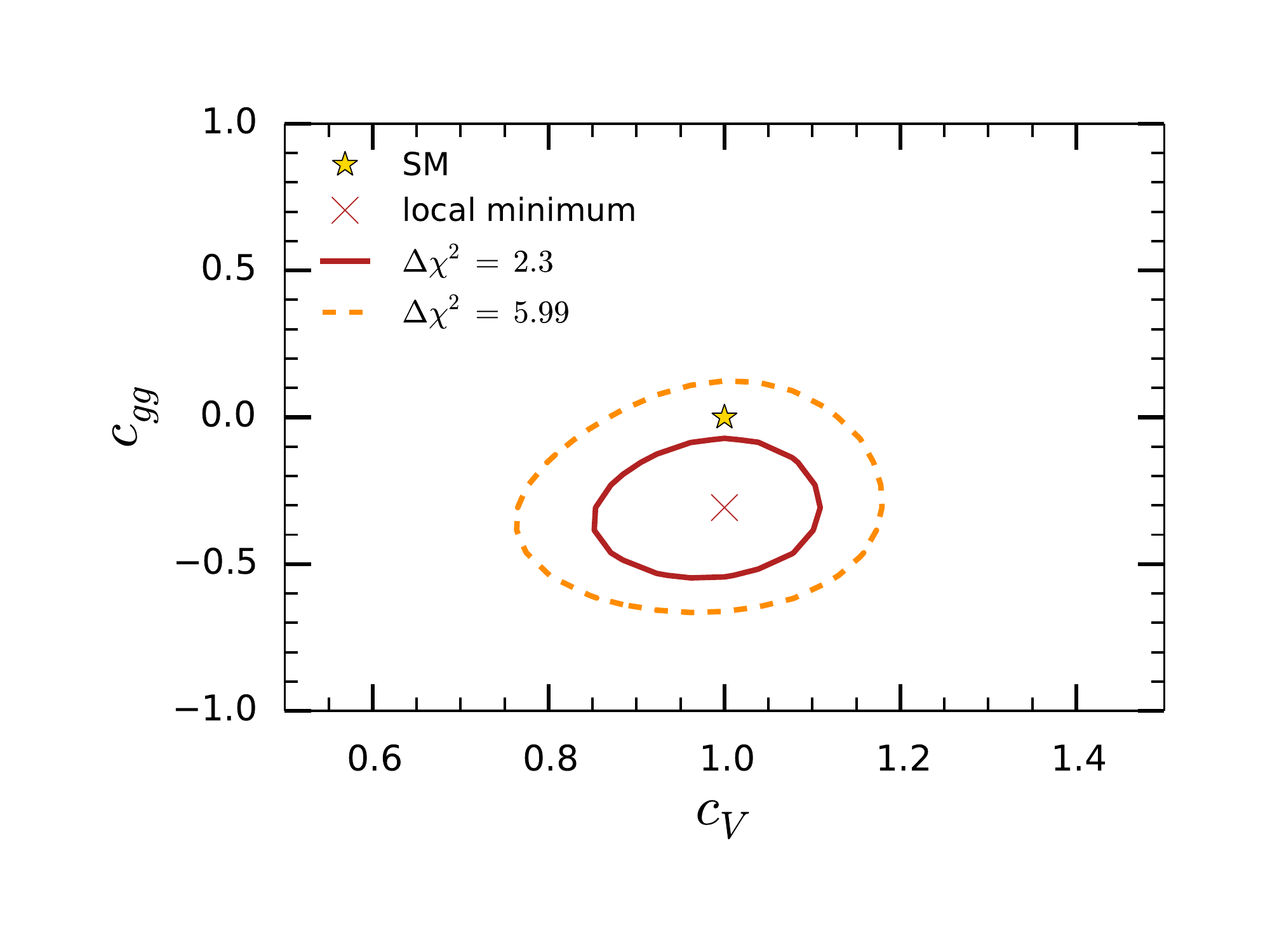}\\ 
\vspace{-0.6cm}
\includegraphics[width=0.4\textwidth]{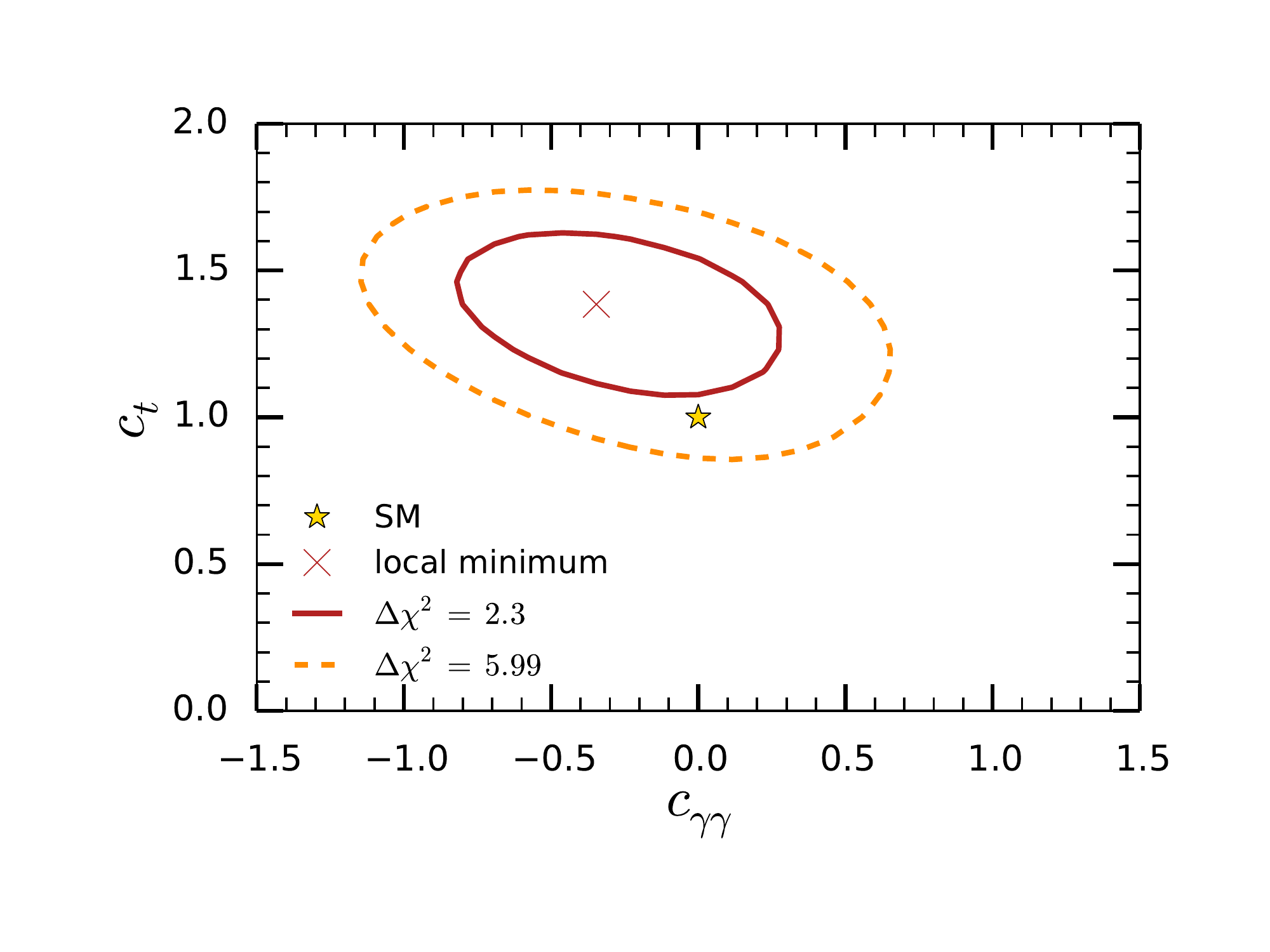}
~
\includegraphics[width = 0.4\textwidth]{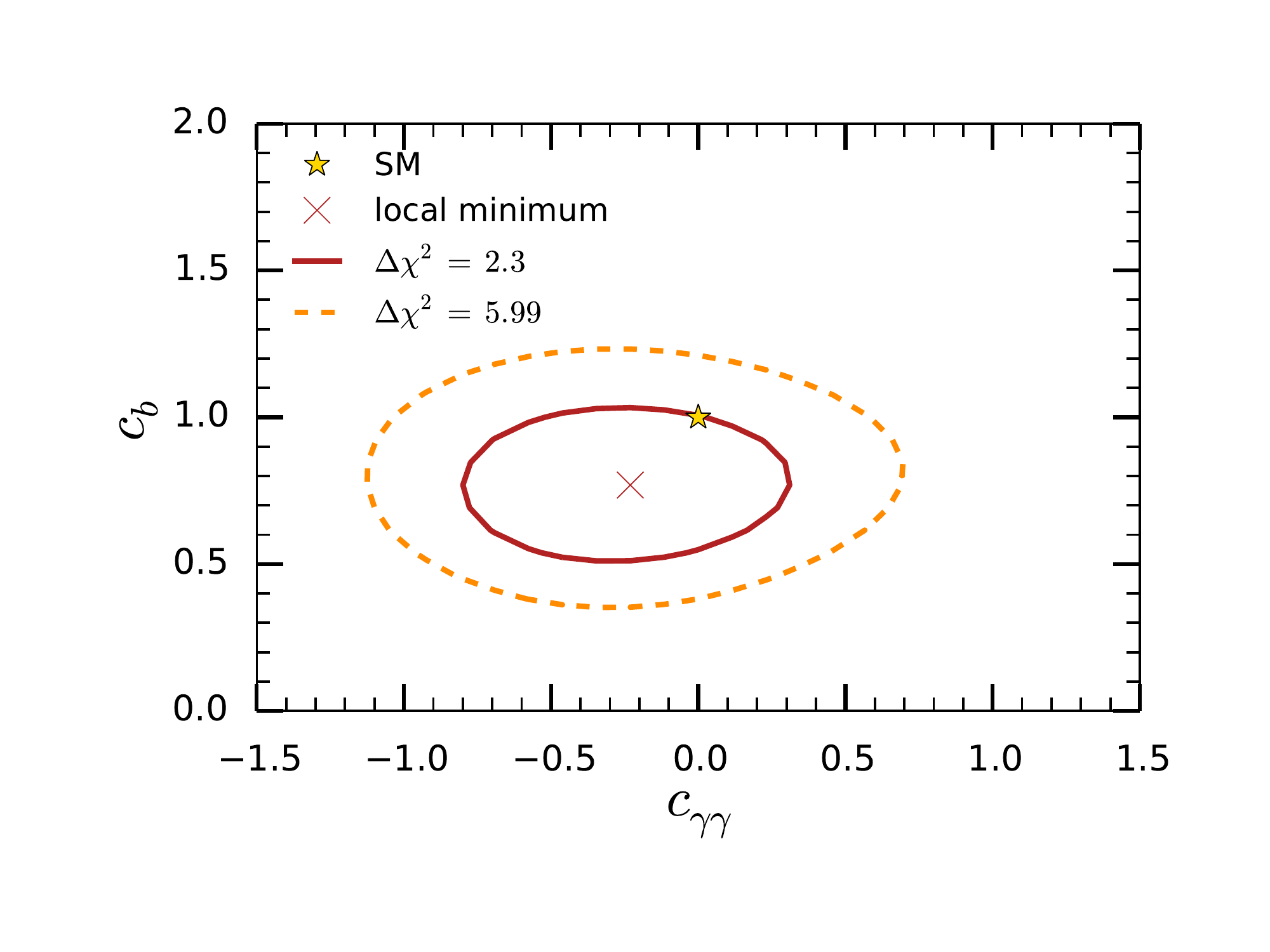}\\ 
\vspace{-0.6cm}
\includegraphics[width=0.4\textwidth]{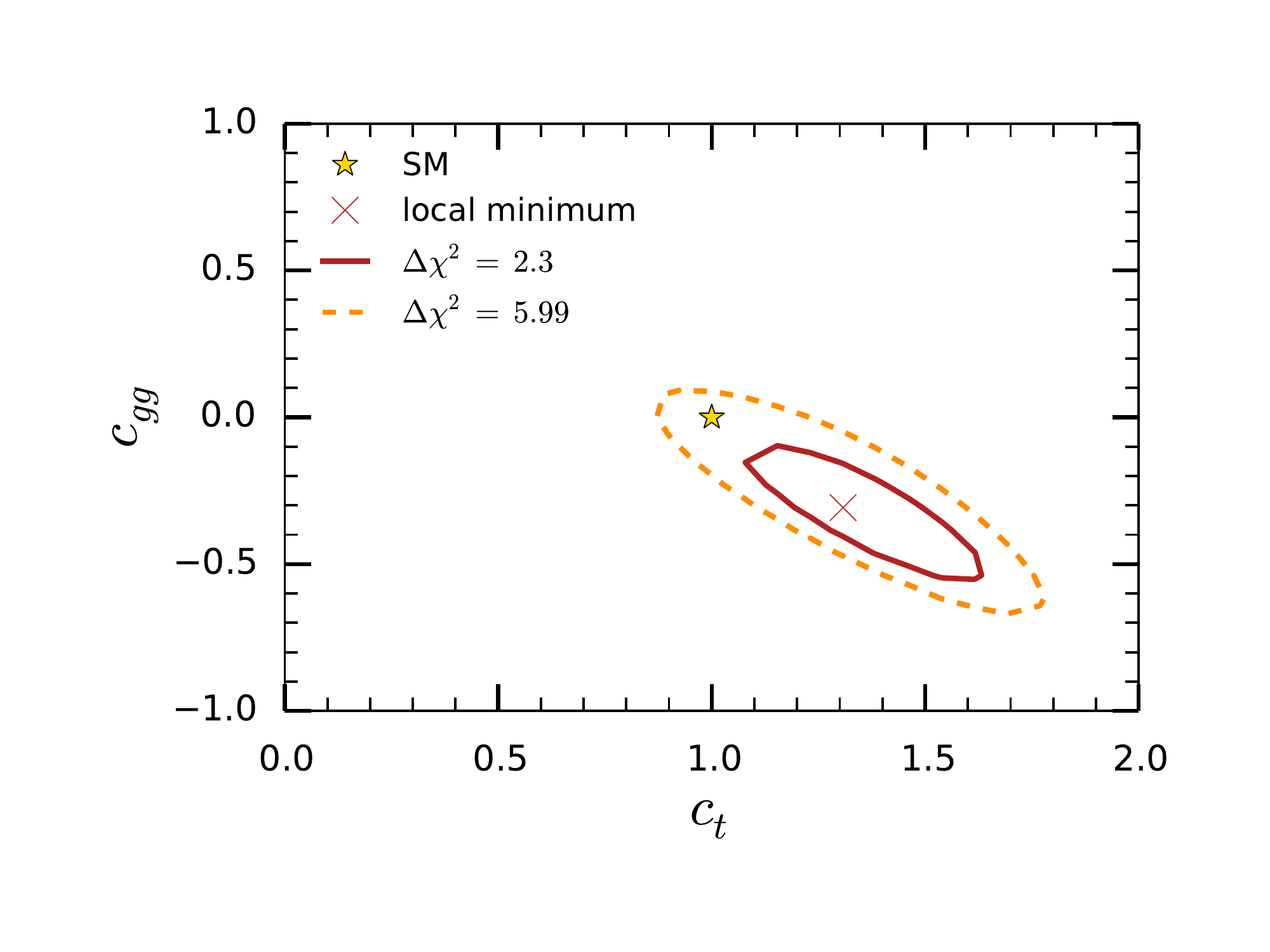}
~
\includegraphics[width = 0.4\textwidth]{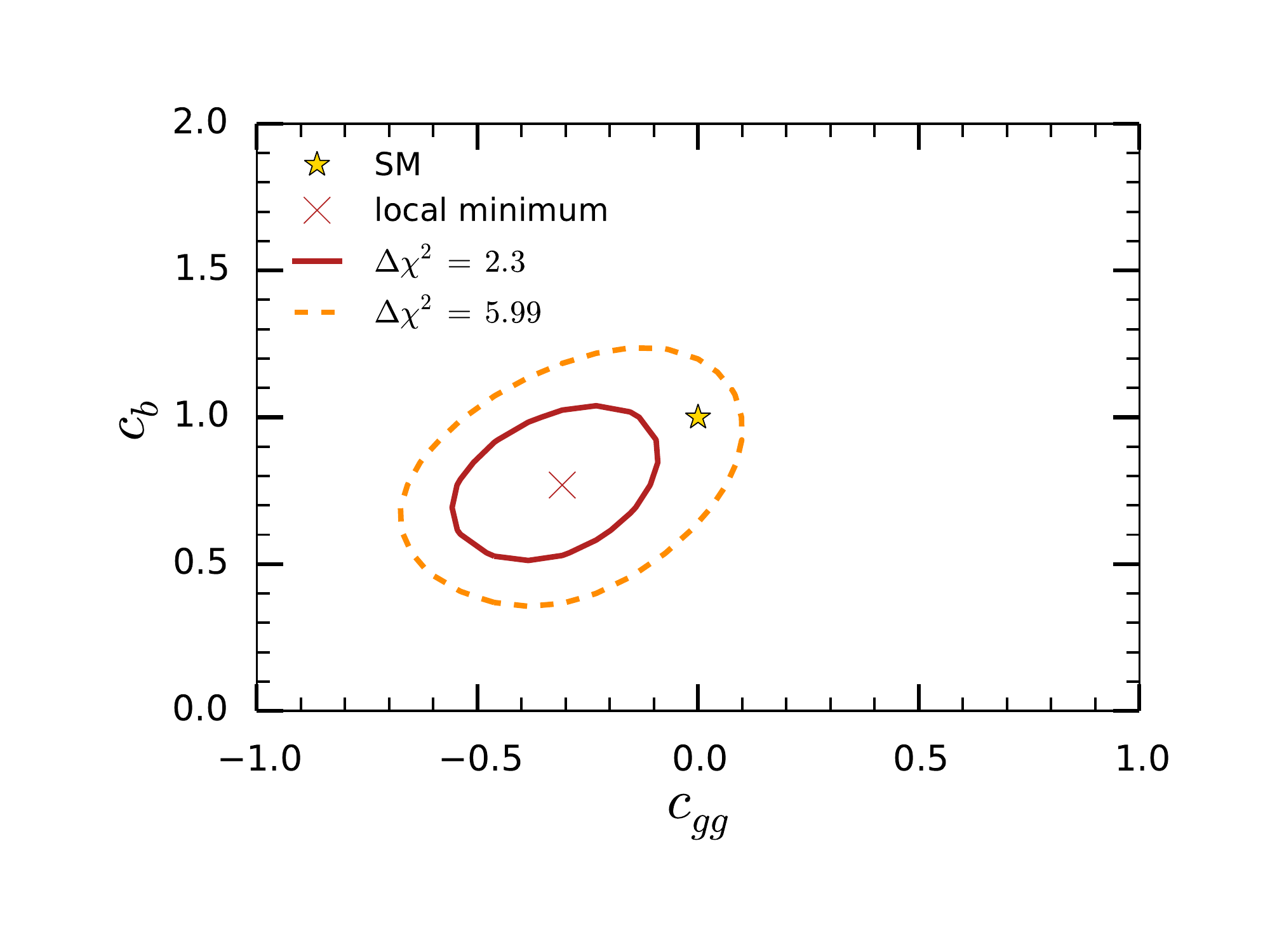}
\vspace{-1cm}
\caption{\label{2dplots} \it \small  $\Delta \chi^2$ isocontours for the 
two-dimensional marginalized posterior pdf.     }
\end{figure}
Figure~\ref{2dplots} shows isocontours of $\Delta \chi^2$ for the 
two-dimensional marginalized pdf for several combinations of parameters.  
Isocontours of $\Delta \chi^2= 2.3, 5.99$ correspond to $68\%, 95\%$ 
Bayesian credible regions to a good accuracy. The minimum of the $\chi^2$ 
(maximum of the marginalized pdf) and the SM point are shown in each case.    
A particularly strong anticorrelation is obtained between $c_{t}$ and $c_{gg}$ 
since the associated contributions to the Higgs production cross-section via 
gluon-fusion interfere constructively and have a similar size, see 
Appendix~\ref{sec:HS}.    Significant correlations are also obtained for 
$c_V - c_b$ and $c_b - c_{\tau}$, as seen in Eq. \eqref{eq:fitresult}.    
The results of the fit show that deviations from the Standard Model are within 
1--2 $\sigma$, which corresponds to an ${\cal O}(10\%)$ uncertainty 
in the Higgs couplings.


The priors used for $\{c_V, c_{t}, c_{b}, c_{\tau}, c_{\gamma \gamma}, c_{gg}\}$ 
in the previous analysis have played the role of uninformative priors, only 
excluding values of the $c_{i}$ that would be unnaturally large within the 
EFT. The posterior pdf is controlled in this case by the likelihood function. 
The relevance of the Bayesian analysis becomes manifest when we address the 
stability of the fit and consider modifications of Higgs couplings for which 
the experimental information is scarce at the moment.   

Our fit can naturally be extended by including modifications of the Higgs 
couplings to light fermions and a local contribution to 
$h \rightarrow Z \gamma$, all of which enter at leading order and should 
therefore be taken into account together with the set 
$\{c_V, c_{t}, c_{b}, c_{\tau}, c_{\gamma \gamma}, c_{gg}\}$. Including in the fit 
modifications of the Higgs coupling to muons $(c_{\mu})$ and a local 
contribution to $h\rightarrow Z \gamma$ $(c_{Z\gamma})$ will not affect the 
joint pdf for the variables 
$\{c_V, c_{t}, c_{b}, c_{\tau}, c_{\gamma \gamma}, c_{gg}\}$ given the current 
experimental bounds from $h \rightarrow \mu \mu$ and 
$h \rightarrow Z \gamma$~\cite{combid}.  The marginal distributions for 
$c_{\mu}$ and $c_{Z\gamma}$ will however be strongly sensitive to the prior 
choice given that the data is not sensitive yet to $\mathcal{O}(10-20\%)$ 
deviations in these couplings. Extending the analysis by considering 
modifications of the additional Higgs couplings to light fermions 
($e, u,d, c, s$) would potentially lead to overfitting and make the fit 
highly unstable on the other hand.  By imposing priors that restrict these 
couplings to be of natural size within the EFT, the stability of the fit is 
recovered and the joint pdf for 
$\{c_V, c_{t}, c_{b}, c_{\tau}, c_{\gamma \gamma}, c_{gg}\}$ remains basically 
unaffected.\footnote{We have verified these points by probing the posterior 
pdf with a Markov Chain Monte Carlo.}  A generic discussion of the use of 
Bayesian priors and the problem of overfitting in EFT parameter estimation 
has been given recently in Ref.~\cite{Wesolowski:2015fqa}.  
 
The naturalness priors on the low-energy constants are also crucial for 
estimating the truncation error associated with higher-order contributions 
in the EFT expansion. The latter can be considered negligible given the 
current precision on the extraction of the leading contributions, 
see Sec.~\ref{sec:NLO} for a discussion of these corrections.


We end this section by commenting on the $\kappa$ formalism adopted by the 
ATLAS and CMS collaborations for the interpretation of Higgs 
data~\cite{combid,Heinemeyer:2013tqa}.\footnote{We refer specifically to the 
$\kappa$ formalism as defined in Sec.~10.3.7 of \cite{Heinemeyer:2013tqa} 
with $\kappa_W = \kappa_Z \equiv \kappa_V$.}  Higgs coupling modifiers are 
defined in the $\kappa$ formalism such that 
$\kappa_j^2 = \sigma_j/\sigma_j^{\mbox{\scriptsize{SM}}}$ ($\kappa_j^2 = 
\Gamma_j/\Gamma_j^{\mbox{\scriptsize{SM}}}$) for a given production process 
(decay mode). In the SM all the $\kappa_j$ are equal to unity by definition.
Considering only third generation fermion Higgs couplings and custodial 
invariance one arrives at a set of six coupling modifiers 
$\{\kappa_V,\kappa_t,\kappa_b,\kappa_{\tau},\kappa_{\gamma},\kappa_{g} \}$
\cite{combid,Heinemeyer:2013tqa}. The individual Higgs coupling modifiers 
$\kappa_{V,t,b,\tau}$ correspond to our parameters $c_{V,t,b,\tau}$. 
An expression for the effective coupling modifiers $\kappa_{g, \gamma}$ in 
terms of our parameters can be read from (\ref{eq:6}) in the Appendix.       

To the best of our knowledge the experimental collaborations have not yet
reported the results of a global fit of 
$\{\kappa_V,\kappa_t,\kappa_b,\kappa_{\tau},\kappa_{\gamma},\kappa_{g} \}$ 
including the full covariance matrix.  We performed a global Bayesian 
inference analysis using the parameters $\kappa_j$ with a flat prior in the 
ranges $\kappa_V\in [0.5,1.5],\kappa_{t,b,\tau}\in [0,2]$ and
$\kappa_{\gamma,g}\in [0.5,1.5]$. 
Neglecting the small absorptive parts of the loop functions 
with light internal fermions in  
$\Gamma(h \rightarrow gg, \gamma \gamma)$, the relation between the 
$\kappa_j$ and the $c_j$ variables amounts to a linear transformation.   
We approximate the posterior pdf obtained for the $\kappa_j$ by a multivariate 
normal distribution. We are then able to recover our results in 
(\ref{eq:fitresult}) to a reasonable accuracy by performing the 
corresponding change of variables.     

Based on the previous analysis, we find recent criticism of the $\kappa$ 
formalism unjustified. As explained in the previous section, the $\kappa$ 
formalism has a solid theoretical interpretation within the electroweak 
chiral Lagrangian \cite{Buchalla:2015wfa}. In no way it
should be considered a mere phenomenological parametrization, with no 
relation to field theory. On the contrary, it is rooted in EFT and it thus 
allows systematic improvements (higher order QCD, electroweak, and new 
physics corrections) to be incorporated.

\section{NLO corrections}
\label{sec:NLO}

\subsection{Nonlinear EFT beyond LO}

The fact that the Lagrangian in (\ref{llopar2}) can be mapped onto the leading 
order chiral Lagrangian (in the unitary gauge) means that (\ref{llopar2}) is 
embedded in a systematic expansion. In particular, this implies that the 
results we presented in the fit of the previous section are accurate up to 
corrections of relative order $\xi/16\pi^2$.
In this section we will discuss how one can include NLO new physics effects systematically in each of the Higgs decay modes. Within the chiral Lagrangian these are corrections of order $\xi/16\pi^2\lesssim 0.1\%$, well beyond the precision levels expected for Higgs couplings at the LHC, even in its final stages. Our discussion is therefore meant to illustrate how the systematics of the expansion works and aimed at eventual future colliders. For illustration we will concentrate on Higgs decays.

Since nonlinear EFTs are based on loop expansions, NLO counterterms and 
one-loop diagrams made of tree level vertices contribute at the same order. 
The full set of operators needed for the Higgs decays $h\to Zl^+l^-$, 
$h\to f\bar f$ or $h\to Z\gamma$ up to NLO is
\begin{align}
\label{Lageff1}
\mathcal{L}&= 
\left(m_W^2 W_{\mu}^{+}W^{- \mu}+\frac{1}{2}m_Z^2 Z_{\mu} Z^{\mu}\right) 
\left(2 c_V \frac{h}{v}+2c_{V2}\frac{h^2}{v^2}\right)  
+\delta_c m^2_Z Z_\mu Z^\mu c_V\frac{h}{v}\nonumber\\
&-c_3\frac{h^3}{v^3}
-\sum_{f} y_f \bar f f \left(c_fh+c_{f2}\frac{h^2}{v}\right)\nonumber\\
&+Z_{\mu}{\bar{\ell}}\gamma^{\mu}\Big[g_V-g_A\gamma_5\Big]\ell+
\frac{h}{v}Z_{\mu}{\bar{\ell}}\gamma^{\mu}\Big[g_{Vh}-g_{Ah}\gamma_5\Big]\ell+
\left(g_W W_{\mu}{\bar{\ell}}\gamma^{\mu}\nu+
g_{Wh}\frac{h}{v}W_{\mu}{\bar{\ell}}\gamma^{\mu}\nu+{\mathrm{h.c.}}\right)
\nonumber\\
&+ \frac{e^2}{16 \pi^2 } c_{\gamma \gamma} F_{\mu\nu} F^{\mu\nu} \frac{h}{v}
+\frac{e g'}{16 \pi^2 } c_{Z \gamma} Z_{\mu\nu} F^{\mu\nu} \frac{h}{v}
+c_{ZZ}\frac{g'^2}{16 \pi^2}Z_{\mu \nu} Z^{\mu\nu}\frac{h}{v}
+c_{WW}\frac{g^2}{16 \pi^2} W_{\mu \nu}^+ W^{-\mu\nu} \frac{h}{v}
\nonumber\\
&+ \frac{g_s^2}{16 \pi^2} c_{gg}\langle G_{\mu \nu} G^{\mu\nu}\rangle\frac{h}{v}\,
\end{align}
which are the relevant operators of the chiral Lagrangian up to NLO in unitary gauge,\footnote{The gauge interactions in the second line are restricted to leptons but can be trivially extended to include quarks.} with $W_{\mu\nu}^{\pm}\equiv\partial_{\mu}W_{\nu}^{\pm}-\partial_{\nu}W_{\mu}^{\pm}$. If custodial symmetry breaking is induced by the weak sector, then the following relation holds: 
\begin{align}\label{cirel}
c_{ZZ}s_W^2-\frac{1}{2}c_{WW}+c_{Z\gamma}s_W^3+c_{\gamma\gamma}s_W^4=0
\end{align}
such that one of the couplings can be expressed in terms of the others. Renormalization of fields and couplings is implicitly assumed, such that the LO Higgs couplings are now of the form    
\begin{align}
c_{V,f}=1+{\cal{O}}(\xi)+{\cal{O}}\left(\xi/16\pi^2\right)
\end{align}  
whereas the gauge couplings are
\begin{align}
g_{V,A,W}=g_{V,A,W}^{(0)}+\delta g_{V,A,W}
\end{align}
where $g_{V,A,W}^{(0)}$ are the SM values and $\delta g_{V,A,W}\sim {\cal{O}}\left(\xi/16\pi^2\right)$ can be computed from the NLO operators of the chiral Lagrangian. Note that while at LO custodial symmetry was preserved, and therefore $c_W=c_Z=c_V$, the inclusion of NLO effects generically breaks custodial symmetry at the per-mille level, in agreement with LEP bounds. The parameter $\delta_c \sim {\cal{O}}(\xi/16\pi^2)$ captures this effect. 

\subsection{Higgs decays at NLO and comparison 
with the linear EFT}\label{sec:4.1}

\begin{figure}[t]
\centering
\includegraphics[width=15.5cm]{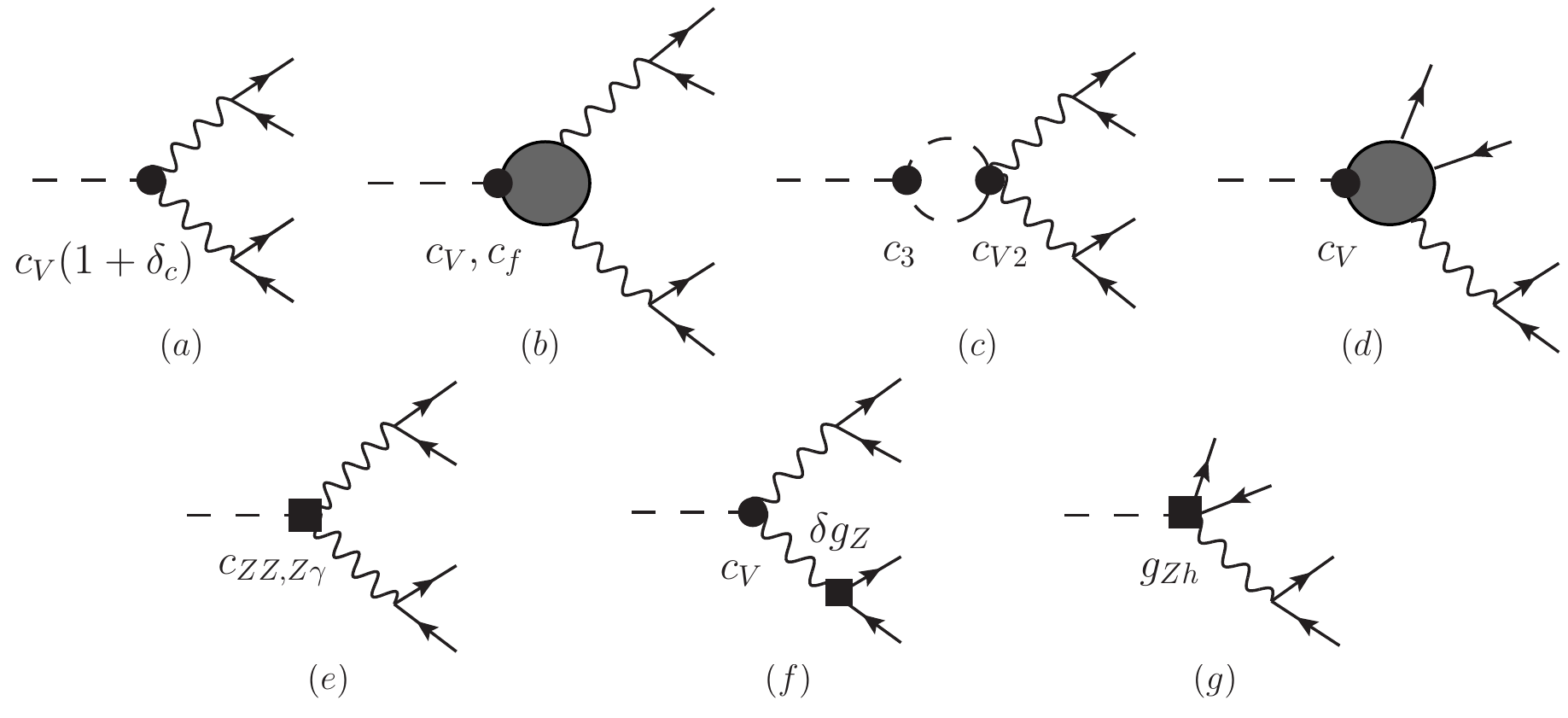}
\caption{\it \small Diagrams contributing to $h\to Zl^+l^-$ up to NLO in the 
nonlinear EFT expansion, ${\cal{O}}(\xi/16\pi^2)$. The blobs are the SM loop 
contributions~\cite{Kniehl:1990mq,CarloniCalame:2006vr,Bredenstein:2006rh}. 
The black circles (squares) are vertices from the LO (NLO) Lagrangian,
where $\delta_c$ is a NLO effect. Pure SM gauge-boson propagator and vertex 
corrections exist for diagram (a), which are not explicitly shown. 
(c) is a representative of loop diagrams with internal Higgs lines. 
The $Z$-boson line in (d) may also be 
attached to the external fermion lines.}\label{hZZ}
\end{figure}
We first consider the process $h\to Zl^+l^-$.
The set of diagrams contributing to this decay is listed in Fig.~\ref{hZZ}. The upper-left diagram is the leading contribution, which contains the SM and ${\cal{O}}(\xi)$ deviations from it. 
NLO corrections consist of (i) SM loops without Higgs internal lines, namely the $W$ and top loop contributions, which are proportional to $c_V$ and $c_f$, respectively (Fig.~\ref{hZZ}(b), \ref{hZZ}(d)); 
(ii) SM loops with Higgs internal lines, a representative of which is depicted in Fig.~\ref{hZZ}(c), proportional to $c_Vc_{V2}$, $c_3c_{V2}$, $c_3c_V^2$ and $c_V^3$; and (iii) NLO counterterms listed in the second row. All NLO contributions are consistently of ${\cal{O}}(\xi/16\pi^2)$. As discussed in~\cite{Buchalla:2013mpa}, a remarkable feature of the nonlinear EFT is that the decay rate is sensitive to LO new physics effects, while differential distributions probe the NLO corrections. Accordingly, while deviations from the SM in the decay rates can be easily expected to reach the $10\%$ level, new physics effects in asymmetries are typically expected at the per-mille level.\footnote{Actually, they happen to be enhanced at the low percent level~\cite{Buchalla:2013mpa} due to the smallness of $g_V$ for the electron.}  

At this point it is instructive to compare with the same process in the EFT with linearly realized EWSB 
\cite{Passarino:2012cb,Einhorn:2013tja,Falkowski:2014tna,Ghezzi:2015vva,Hartmann:2015oia}.
In this case new physics effects enter at NLO 
(dimension 6) and are proportional to $v^2/\Lambda^2\equiv \varepsilon$. If $\Lambda\sim 1$ TeV, then $\varepsilon> (16\pi^2)^{-1}$ and new physics contributions are bigger than SM loop effects. A larger $\Lambda$ ($\gtrsim 3$ TeV) spoils this numerical hierarchy, while a smaller $\Lambda$ ($< 1$ TeV) jeopardizes the convergence of the EFT expansion and might eventually be in conflict with exclusion limits. Most of the studies with the linear EFT are done assuming, implicitly or explicitly, this fiducial window for the new physics scale $\Lambda$.

In~\cite{Arzt:1994gp} it was argued that, if UV completions are assumed to be weakly-coupled and renormalizable, a loop counting on the NLO operator basis can be applied on top of pure dimensional power-counting. Then NLO operators that can be tree-level generated in a UV completion are generically more relevant than the ones that can only be loop generated, which can be neglected. For $h\to Zl^+l^-$, the argument amounts to dropping the diagram in Fig.~\ref{hZZ}(e), of order $\varepsilon/16\pi^2$, while keeping diagrams \ref{hZZ}(f) and \ref{hZZ}(g), of order $\varepsilon$. This is the approach taken for instance in~\cite{Ghezzi:2015vva}. In some analyses it is further argued that these remaining contributions can be dropped based on LEP constraints \cite{Gupta:2014rxa}. 
Leaving aside how legitimate this assumption might be,\footnote{See e.g.~\cite{Buchalla:2013mpa} for a discussion on why dropping the $hZ\ell^+\ell^-$ diagram is not justified and~\cite{Berthier:2015gja} for a more general discussion on how LEP precision constraints translate into constraints on EFT coefficients.} if gauge corrections are assumed to be suppressed, the dominant new physics effects are contained in the shift contributions to $c_V$,
\begin{align}\label{cve}
c_V=1+{\cal{O}}(\varepsilon)
\end{align}
\begin{figure}[t]
\centering
\includegraphics[width=8.5cm]{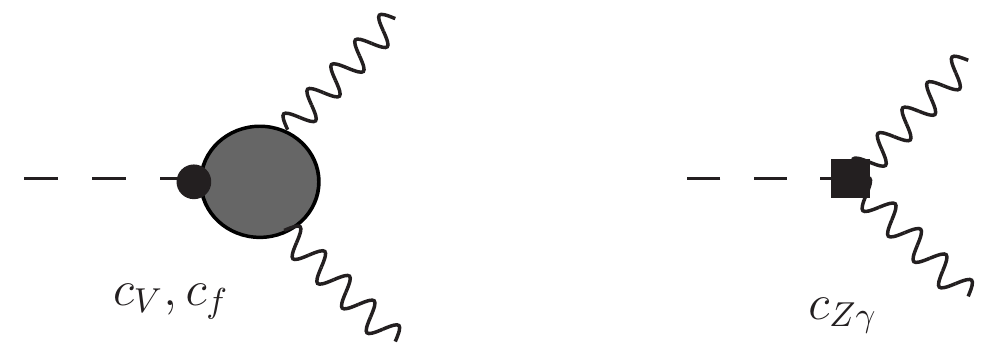}
\caption{\it \small Diagrams contributing to $h\to Z\gamma$ up to NLO in the nonlinear expansion, ${\cal{O}}(\xi/16\pi^2)$.}\label{hZg}
\end{figure}
It is important to stress that the simple picture that comes out of 
(\ref{cve}) follows from adopting dimensional counting supplemented by 
a number of additional assumptions, namely
\begin{description}
\item[(a)] dimension six operators dominate over Standard Model loops, 
$\varepsilon > (16\pi^2)^{-1}$;
\item[(b)] a (UV-based) loop counting is added on top of the (IR-based) 
power-counting;
\item[(c)] there are additional suppressions based on 
phenomenologically-motivated considerations.
\end{description}
The overall effect of (a), (b) and (c) is to generate additional hierachies not present in the EFT power-counting: new physics effects over SM loops and Higgs observables over LEP-probed ones. Internal consistency of the EFT in any case limits the new physics effects to be around the few percent level. 
In contrast, in the nonlinear case, 
(i) the different hierarchies are dynamically imprinted in the power-counting: corrections in the gauge sector are expected at $\xi/16\pi^2$ and the current experimental pattern of 1-2 orders of magnitude between Higgs and electroweak precision is realized parametrically;
and (ii) the new physics effects can naturally accommodate ${\cal{O}}(10\%)$ deviations in the Higgs sector without jeopardizing the convergence of the expansion.\footnote{Notice that $\varepsilon\sim \Lambda^{-2}$ is necessarily linked to new physics thresholds, while $\xi\sim f^{-2}$ is just a symmetry breaking scale. In this context, it is interesting to note that a scenario where the scale $f$ is populated by new physics states~\cite{Falkowski:2007hz,Carena:2014ria} is within the applicability range of the nonlinear EFT, provided $f$ is sufficiently large with respect to $v$. This scenario has a number of phenomenologically interesting aspects, some of which have been discussed in a previous paper~\cite{Buchalla:2014eca}.}

It is worth emphasizing that in the nonlinear EFT $c_W=c_Z=c_V$ holds to leading chiral order, i.e. to all orders in $\xi$, while custodial symmetry breaking terms will in general break this degeneracy at NL chiral order, 
$c_W-c_Z\sim \xi/16\pi^2$. Within the linear EFT $c_W-c_Z\sim \varepsilon$ unless one assumes custodial symmetry breaking to be numerically small.

Let us now turn our attention to loop-induced processes. 
In this case we restrict our consideration to the level of
${\cal O}(\xi/16\pi^2)$ corrections. These are NLO terms in view of the global
loop counting, although they are only leading-order effects of ${\cal O}(\xi)$
relative to the SM one-loop amplitude.
Consider for instance the leading contributions to $h\to Z\gamma$ within the nonlinear EFT, which are summarized in Fig.~\ref{hZg}. The first diagram collects the SM $W$ and top quark loop contributions multiplied respectively by $c_V$ and $c_f$ and the second one the NLO counterterm. Since $c_{V,f}\sim 1+{\cal{O}}(\xi)$ and $c_{Z\gamma}\sim \xi/16\pi^2$, the new-physics piece is consistently of order $\xi/16\pi^2$. In the linear EFT with the additional assumptions 
(a) and (b) mentioned above one finds the same topologies, with all contributions homogeneously of order $\varepsilon/16\pi^2$. Again, if ${\cal{O}}(\varepsilon)$ corrections to the $Z$ couplings to fermions are dropped using phenomenological arguments, the leading order new physics effects enter as a shift effect on 
$c_{V,f}=1+{\cal{O}}(\varepsilon)$ and through the local term
$c_{Z\gamma}\sim\varepsilon/16\pi^2$. This is formally similar to the nonlinear
EFT, however only at the price of additional assumptions. 
It is interesting to note that in $h\to\gamma\gamma$ gauge corrections are absent altogether because of electroweak gauge invariance, and there is no need to resort to LEP bounds. 

In comparing the linear and nonlinear EFT parametrizations in loop-induced processes, one should keep in mind that in the nonlinear case the new physics corrections appear already at LO, of order $\xi/16\pi^2$, while in the linear EFT they are a NLO effect, of order $\varepsilon/16\pi^2$. Moreover, the previous considerations only apply if the scheme suggested in~\cite{Einhorn:2013tja} 
is adopted, corresponding to the assumptions (a) -- (c) above. 
If pure dimensional counting is employed in the linear EFT without additional assumptions, the number of diagrams contributing to $h\to \gamma\gamma$ 
or $h\to Z\gamma$ at leading order in new physics corrections is substantially larger and the global picture gets more complicated~\cite{Hartmann:2015oia}.
Consequences for the connection with the conventional $\kappa$ formalism
are discussed below.

\begin{figure}[t]
\centering
\includegraphics[width=13.5cm]{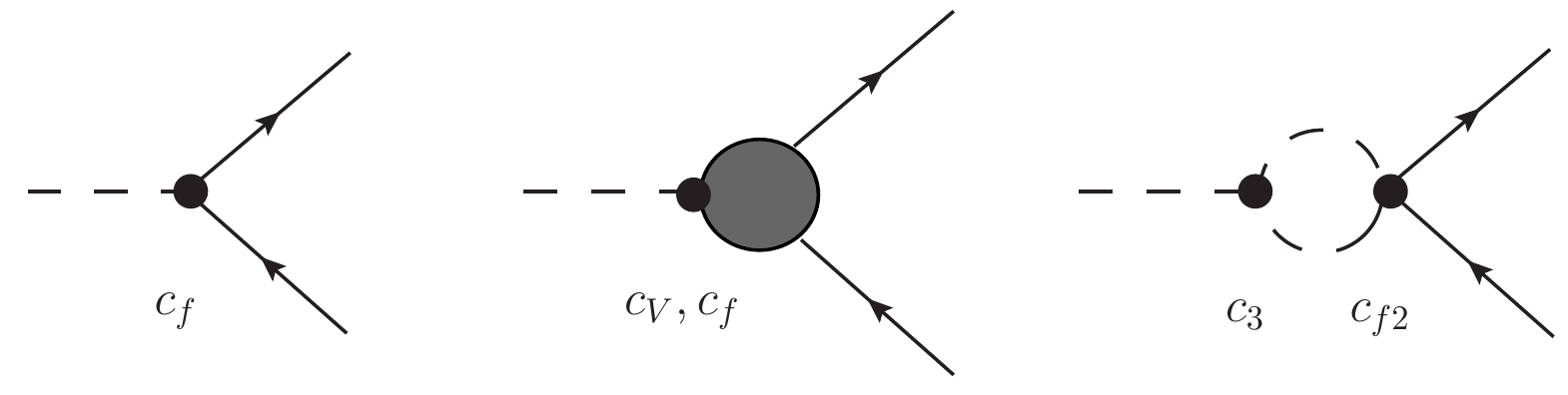}
\caption{\it \small Diagrams contributing to $h\to {\bar{f}}f$ to order 
$\xi/16\pi^2$. The left diagram contains the LO contribution, the central 
diagram shows some of the SM loop topologies 
and the right diagram the genuine chiral loops ($c_f^3$, $c_f c_{f2}$ and 
$c_3c_f^2$ contributions are also to be taken into account).}\label{hff}
\end{figure}

Let us finally comment on $h\to {\bar{f}}f$. The relevant diagrams are collected in Fig.~\ref{hff}. The leading order new physics corrections of order $\xi$ stem from the first diagram. NLO corrections can be divided into (i) SM-like topologies without Higgs internal lines (second diagram), with contributions of order $1/(16\pi^2)(1+{\cal{O}}(\xi))$ and (ii) diagrams with Higgs internal lines, proportional to $c_f^3$, $c_f c_{f2}$, $c_3c_f^2$ and $c_3c_{f2}$. 
The latter is a genuine nonlinear contribution of order $\xi/16\pi^2$. A local counterterm is absent. Within the linear EFT framework of~\cite{Arzt:1994gp}, the main contribution comes from a local dimension-6 operator, which can be absorbed in an effective vertex with coupling $c_f=1+{\cal{O}}(\varepsilon)$~\cite{Einhorn:2013tja}. The leading new physics corrections are again expected at the few percent level.

\subsection{Conventional $\mathbf{\kappa}$ formalism 
as limit of an EFT description}  

An interesting observation can be made on the relation between the EFT 
formulations and the conventional $\kappa$ formalism.
While the nonlinear EFT reproduces, and therefore justifies,
the phenomenological $\kappa$ formalism at leading order in the chiral
expansion \cite{Buchalla:2015wfa}, there is no parametric limit in which 
this is the case for the linear EFT at the level of dimension-6 corrections.
The decay $h\to Z\gamma$ in Fig. \ref{hZg} may serve as an example.
If dimension-6 insertions in the loop diagrams are retained, corrections
unrelated to LO Higgs couplings, e.g. from $\bar ttZ$, are also present.
If these insertions are neglected, only the contact term modifies the SM.
If the contact term is assumed to be loop suppressed, of order
$\varepsilon/16\pi^2$, it has to be dropped as well in the same approximation.  
None of these cases reproduces the conventional $\kappa$ framework. Similar 
comments apply to $h\to Zl^+l^-$. A related discussion of the linear EFT and 
its connection with the $\kappa$ framework has been given in 
\cite{David:2015waa}.

\subsection{QCD loops vs. LO nonlinear EFT}

In general, the systematics of the nonlinear EFT dictates that
one-loop diagrams with vertices from the LO Lagrangian come at the
same order as the NLO local terms. This is particularly true for the
Higgs and electroweak sector, where the inclusion of loop effects
beyond a LO description requires the simultaneous consideration of 
NLO operators, thus increasing the number of free parameters.

On the other hand, it is possible to keep a LO treatment of Higgs couplings
and still consistently include higher-order QCD radiative corrections.
This is because the LO Lagrangian (\ref{l2}), even in the form
(\ref{llopar2}) with the effective $h\to gg$ coupling, is renormalizable
under QCD. Also, the expansion in the QCD coupling is parametrically different
from the chiral expansion and can be considered separately.
This feature is useful in practice, since large radiative effects from QCD
can be taken into account, while otherwise working at LO in the nonlinear EFT.
An interesting example is provided by the discussion of double-Higgs production
in gluon-gluon fusion, $gg\to hh$ in \cite{Grober:2015cwa}, where
anomalous couplings are treated at LO in the nonlinear EFT, but higher-order
QCD corrections are also included.

\subsection{Pseudo-observables}

Pseudo-observables (POs) have been proposed in order to provide a general and 
model-independent link between experimental data and theoretical predictions. 
The main strategy is to identify, at the amplitude level, the 
most general set of independent parameters for each physical process based
on a multiple-pole expansion. A crucial assumption in the PO analysis,
similar to the EFT, is that there are no light undetected particles, 
i.e. a mass gap exists between the electroweak and the TeV scale. 
As of this writing, the pseudo-observable program has only been developed for 
Higgs decays~\cite{Gonzalez-Alonso:2014eva,Ghezzi:2015vva}, working by analogy 
with what was done at LEP for Z-pole observables~\cite{Bardin:1999gt}. 

By construction, the identification of pseudo-observables requires only 
kinematical considerations, leaving the dynamics unspecified. 
In order to interpret the values for the pseudo-observables one needs to 
resort to a dynamical scheme, be it a model or EFTs. 
The nonlinear EFT has the features and advantages discussed in 
\cite{Buchalla:2015wfa} and in the present paper.
It is clear that any PO can be expressed within this EFT in terms
of its parameters. In particular, the expected size of the new-physics 
impact on the PO can be predicted based on the EFT power counting.
For example, the decay rate for $h\to Zl^+l^-$ has been considered as
a PO for $h\to 4l$ in \cite{David:2015waa}. An analysis of this PO
within the nonlinear EFT can be found in \cite{Buchalla:2013mpa}.

\section{Conclusions}
\label{sec:concl}

The main results of this paper can be summarized as follows:

\begin{itemize}
\item 
We have reviewed the electroweak chiral Lagrangian as a
consistent EFT framework to describe new-physics effects
at electroweak energies in a model-independent way.
The emphasis has been on the leading-order (LO) approximation of the
nonlinear EFT, which is equivalent to the conventional 
$\kappa$ formalism. The latter thus receives a proper quantum-field theory 
justification.
\item
The main benefits of the nonlinear EFT at LO are:
(i) It allows one to focus systematically on anomalous couplings of the 
Higgs particle, which could potentially exhibit the largest new-physics 
effects in the electroweak sector. (ii) The limited number of parameters
(as opposed to the full set of dimension-6 corrections) is of considerable
practical importance and will facilitate the interpretation of the data.
(iii) The LO approximation (in new-physics effects) is well adapted to the 
precision foreseen for LHC Run 2.
\item
Concentrating on the processes of Higgs production and decay that
have been measured so far, six parameters of the leading-order EFT
describing anomalous Higgs couplings, are relevant:
$c_V$, $c_t$, $c_b$, $c_\tau$, $c_{\gamma\gamma}$, $c_{gg}$.
Using the {\tt Lilith} code, a fit of these parameters has been performed
to current data within a Bayesian approach. The results agree with the SM
to within $10$--$20\%$. The detailed fit results can be found in
Section \ref{sec:fit}. The new aspect of our analysis is that it is
based on a systematic EFT interpretation of the fit parameters.
\item
We have shown how the LO parametrization can be generalized to
the NLO of the nonlinear EFT. Additional parameters appear at this level,
which however are subleading according to the EFT power counting.
The systematics has been illustrated through various Higgs decays such as
$h\to Zl^+l^-$ or $h\to f\bar f$. The differences with the 
case of the linear EFT including operators of dimension six have also
been discussed.
\end{itemize}

Further important processes that will become accessible in the future, 
such as $h\to Z\gamma$ or double-Higgs production, can be analyzed in the
same way, based on the LO nonlinear EFT, at the expense of introducing
a (small) number of additional couplings.
Our analysis emphasizes the fact that the conventional $\kappa$
framework has a firm foundation as the leading-order approximation
of the nonlinear EFT of the physics at the Terascale.
It will therefore continue to be a powerful and systematic tool
to analyze the physics of the Higgs boson at Run 2 of the LHC
and beyond. 

\section*{Acknowledgements}

We thank Frank Tackmann and Juan Jos\'e Sanz-Cillero for useful
discussions.
This work was performed in the context of the ERC Advanced Grant 
project FLAVOUR (267104) and was supported in part by the DFG grant 
BU 1391/2-1, the DFG cluster of excellence EXC 153 'Origin and Structure of 
the Universe' and the Munich Institute for Astro- and Particle Physics (MIAPP). 
A.C. is supported by a Research Fellowship
of the Alexander von Humboldt Foundation.

\appendix
\numberwithin{equation}{section} 
\section{Higgs signal strengths} 
\label{sec:HS}

The experimental collaborations have provided on-shell Higgs data in the form 
of signal strengths.  These are defined for the different production 
$X \in \{  ggF, WH/ZH, VBF, ttH \}$ and decay channels 
$Y \in \{bb, \tau \tau, WW, ZZ, \gamma \gamma\}$ as
\begin{equation}
\mu =\frac{\sigma(X)\times\mathrm{Br}(h \rightarrow Y)}{
\sigma(X)_{\mbox{\scriptsize{SM}}} 
\times \mathrm{Br}(h\rightarrow Y)_{\mbox{\scriptsize{SM}}} } 
\end{equation}
The leading modifications of the Higgs properties within the nonlinear 
effective theory, encoded in (\ref{llopar2}), amount to a rescaling of the 
relevant Higgs-production cross sections and partial decay widths. 
The relevant Higgs-production cross sections are given by
\begin{align}
\frac{\sigma(\mbox{VH})}{\sigma(\mbox{VH})_{\mbox{\scriptsize{SM}}}} &= c_V^2 
\qquad  \qquad \qquad 
\frac{\sigma(\mbox{VBF})}{\sigma(\mbox{VBF})_{\mbox{\scriptsize{SM}}}} = c_V^2  
\nonumber \\
\frac{\sigma(\mbox{ttH})}{\sigma(\mbox{ttH})_{\mbox{\scriptsize{SM}}}} &=  c_{t}^2 
 \qquad  \qquad \qquad 
 \frac{\sigma(\mbox{ggF})}{\sigma(\mbox{ggF})_{\mbox{\scriptsize{SM}}}}\simeq    
\frac{\Gamma(h\rightarrow gg)}{\Gamma(h\rightarrow gg)_{\mbox{\scriptsize{SM}}}} 
\end{align}
The ratio of the branching ratios can be expressed as
\begin{equation}
  \label{eq:1}
  \frac{\mathrm{Br}(h \rightarrow Y)}{
\mathrm{Br}(h\rightarrow Y)_{\mbox{\scriptsize{SM}}}} = 
\frac{\Gamma^{Y}/\Gamma^{Y}_{\mbox{\scriptsize{SM}}}}{
\sum_{j} (\mathrm{Br}(h\rightarrow j)_{\mbox{\scriptsize{SM}}} 
\times \Gamma^{j}/\Gamma^{j}_{\mbox{\scriptsize{SM}}})}
\end{equation}
The branching ratios of the SM are taken from \cite{Heinemeyer:2013tqa} for 
$m_{h}= 125~\text{GeV}$ and are given to the highest available order in QCD, 
see \cite{Dittmaier:2012vm}. The tree-level decay rates for 
$h \rightarrow VV^*$ $(VV=W^+W^-, ZZ)$ and $h \rightarrow f \bar f $ get 
rescaled compared with the SM by $c_V^2$ and $c_f^2$, respectively. 
For the loop-induced decays \cite{Manohar:2006gz}
\begin{align}
  \label{eq:6}
  \frac{\Gamma(h \rightarrow \gamma \gamma)}{
\Gamma(h \rightarrow \gamma \gamma)_{\mbox{\scriptsize{SM}}}} &=    
\frac{  \left|  \sum_{q}   \frac{4}{3} N_C  Q_q^2    c_{q} A_{1/2}(x_q) 
\eta^{q,\gamma\gamma}_{\text{QCD}}   + \frac{4}{3} c_{\tau}  A_{1/2}(x_{\tau})    + 
c_V A_{1}(x_W) + 2 c_{\gamma \gamma}  \right|^2    }{ \left| \sum_{q}   
\frac{4}{3} N_C  Q_q^2 A_{1/2}(x_q) \eta^{q,\gamma\gamma}_{\text{QCD}}   + 
\frac{4}{3}  A_{1/2}(x_{\tau}) +  A_{1}(x_W)   \right|^2    }  \nonumber \\
 \frac{\Gamma(h \rightarrow gg)}{\Gamma(h \rightarrow gg)_{\mbox{\scriptsize{SM}}}}  
&=  \frac{   \left|  \sum_{q}  \frac{1}{3} c_q A_{1/2}(x_q) \eta^{q,gg}_{\text{QCD}}  
+ \frac{1}{2}  c_{gg}   \right|^2  }{ \left| \sum_{q}  \frac{1}{3}  A_{1/2}(x_q) 
\eta^{q,gg}_{\text{QCD}}    \right|^2    }  
\end{align}
with $x_q =  4 m_q^2/m_h^2$. 
The loop functions are defined as \cite{Gunion:1989we,Djouadi:2005gi}
\begin{align}
    \label{eq:7}
    A_{1/2}(x) &=   \frac{3}{2}  x \left[    1 + (1-x) f(x) \right]    
\qquad  \qquad 
    A_{1}(x)  =  - [  2 + 3 x + 3 x (2-x) f(x)   ] 
\end{align}
 Here
\begin{equation}
\label{eq:8}
f(x)\; =\; \begin{cases} \arcsin^2(1/\sqrt{x})\, , \quad & x\geqslant1 \\[3pt]  
- \dfrac{1}{4} \Big[\ln\Big( \frac{1+\sqrt{1-x}}{1-\sqrt{1-x}}\Big)- 
i\pi \Big]^2\, , & x<1 \end{cases}
\end{equation}
We take into account $\mathcal{O}(\alpha_s)$ corrections due to the exchange 
of hard gluons and quarks in production and decay. To this order, the QCD 
corrections factorize for tree-level amplitudes of production and decay and 
therefore cancel in the ratios.  They do not cancel for 
$h \rightarrow \gamma \gamma, gg$, where we included 
$\eta^{t,gg}_{\text{QCD}}  = 1+11\alpha_{s}/4\pi$ and 
$\eta^{t,\gamma\gamma}_{\text{QCD}} = 1 -\alpha_{s}/\pi$ for the 
top loop \cite{Manohar:2006gz,Gunion:1989we,Djouadi:2005gi,Contino:2013kra,Contino:2014aaa}. 
The effects of QCD corrections on other quark loops were checked to be 
negligibly small. We have neglected non-factorizing electroweak corrections, 
which are expected to be small.


\end{document}